\documentclass[aps,pre,reprint,groupedaddress]{revtex4-2}

\usepackage{graphicx}
\usepackage{bm}
\usepackage{amssymb}
\usepackage{amsbsy}
\usepackage{amsmath}
\usepackage{mathtools}
\usepackage{xcolor}

\usepackage{color}
\usepackage{tikz}
\usetikzlibrary{shapes}
\usepackage{soul}

\begin{document}


\title{Non-Brownian diffusion and chaotic rheology of autophoretic disks}


\author{R. Kailasham}
\affiliation{Department of Chemical Engineering, Carnegie Mellon University, Pittsburgh, PA 15213, USA}
\author{Aditya S. Khair}
\email{akhair@andrew.cmu.edu}
\affiliation{Department of Chemical Engineering, Carnegie Mellon University, Pittsburgh, PA 15213, USA}


\date{\today}

\begin{abstract}
The dynamics of a two dimensional autophoretic disk is quantified as a minimal model for the chaotic trajectories undertaken by active droplets. Via direct numerical simulations, we show that the mean-square displacement of the disk in a quiescent fluid is linear at long times. Surprisingly, however, this apparently diffusive behavior is non-Brownian, owing to strong cross-correlations in the displacement tensor. The effect of a shear flow field on the chaotic motion of an autophoretic disk is examined. Here, the stresslet on the disk is chaotic for weak shear flows; a dilute suspension of such disks would exhibit a chaotic shear rheology. This chaotic rheology is quenched first into a periodic state and ultimately a steady state as the flow strength is increased.
\end{abstract}


\makeatletter
\def\squarecorner#1{
    %
    \pgf@x=\the\wd\pgfnodeparttextbox%
    \pgfmathsetlength\pgf@xc{\pgfkeysvalueof{/pgf/inner xsep}}%
    \advance\pgf@x by 2\pgf@xc%
    \pgfmathsetlength\pgf@xb{\pgfkeysvalueof{/pgf/minimum width}}%
    \ifdim\pgf@x<\pgf@xb%
        \pgf@x=\pgf@xb%
    \fi%
    %
    \pgf@y=\ht\pgfnodeparttextbox%
    \advance\pgf@y by\dp\pgfnodeparttextbox%
    \pgfmathsetlength\pgf@yc{\pgfkeysvalueof{/pgf/inner ysep}}%
    \advance\pgf@y by 2\pgf@yc%
    \pgfmathsetlength\pgf@yb{\pgfkeysvalueof{/pgf/minimum height}}%
    \ifdim\pgf@y<\pgf@yb%
        \pgf@y=\pgf@yb%
    \fi%
    %
    \ifdim\pgf@x<\pgf@y%
        \pgf@x=\pgf@y%
    \else
        \pgf@y=\pgf@x%
    \fi
    %
    \pgf@x=#1.5\pgf@x%
    \advance\pgf@x by.5\wd\pgfnodeparttextbox%
    \pgfmathsetlength\pgf@xa{\pgfkeysvalueof{/pgf/outer xsep}}%
    \advance\pgf@x by#1\pgf@xa%
    \pgf@y=#1.5\pgf@y%
    \advance\pgf@y by-.5\dp\pgfnodeparttextbox%
    \advance\pgf@y by.5\ht\pgfnodeparttextbox%
    \pgfmathsetlength\pgf@ya{\pgfkeysvalueof{/pgf/outer ysep}}%
    \advance\pgf@y by#1\pgf@ya%
}
\makeatother

\pgfdeclareshape{square}{
    \savedanchor\northeast{\squarecorner{}}
    \savedanchor\southwest{\squarecorner{-}}

    \foreach \x in {east,west} \foreach \y in {north,mid,base,south} {
        \inheritanchor[from=rectangle]{\y\space\x}
    }
    \foreach \x in {east,west,north,mid,base,south,center,text} {
        \inheritanchor[from=rectangle]{\x}
    }
    \inheritanchorborder[from=rectangle]
    \inheritbackgroundpath[from=rectangle]
}

\DeclareRobustCommand{\markerone}{\raisebox{0pt}{\tikz{\node[draw=red,scale=0.4,circle,fill=none](){};}}}
\DeclareRobustCommand{\markertwo}{\raisebox{0pt}{\tikz{\node[draw=black,scale=0.4,square,fill=none](){};}}}
\DeclareRobustCommand{\markerthree}{\raisebox{0pt}{\tikz{\node[draw=black!60!green,scale=0.4,isosceles triangle,rotate=90,fill=black!60!green](){};}}}

\DeclareRobustCommand{\markerfive}{\raisebox{0pt}{\tikz{\node[draw=blue,scale=0.4,diamond,fill=none](){};}}}

\def\tr{\textcolor{red}}
\def\tb{\textcolor{blue}} 

\maketitle


\section{\label{sec:intro}Introduction}

{An oil (or water) droplet immersed in a surfactant solution above the critical micelle concentration isotropically emits swollen micelles from its surface. The diffusiophoretic interaction between the droplet and its products of solubilization could result in spontaneous self-propulsion of the droplet, through a symmetry-breaking instability~\cite{Izri2014,Maass2016,Suda2021,Decayeux2021,Hu2022,Michelin2022b}. This instability occurs when the Peclet number ($Pe$), i.e., the dimensionless ratio of the strength of advective to diffusive transport of the products of solubilization, exceeds a critical value. Such active droplets display varying patterns of motion, including straight, curvilinear, or meandering, as $Pe$ is increased, before entering a chaotic regime at high enough values of the P\'{e}clet number~\cite{Suga2018,Hokmabad2021}.} The occurrence of chaotic dynamics~\cite{Muller2018,Muller-Bender2023} is intriguing since the fluid flow around the drop is at low Reynolds number, where nonlinear inertial forces are absent.

In this article, we consider a two-dimensional autophoretic disk as a minimal model for an active drop {(see Supplementary Material~\cite{FNote1} for the rationale behind this representation)} and discover two important features about its dynamics. First, although such particles undergo normal diffusion at long time scales~\cite{Hu2019}, we show that they are not Brownian due to strong cross-correlations in their displacement components. Second, we calculate the motion of an autophoretic disk in a shear flow field. Here, the velocity and stresslet of the disk are chaotic at low values of the shear rate, implying that a dilute suspension of disks would exhibit a chaotic shear rheology. 
On increasing the strength of the shear flow (at fixed P\'{e}clet number) the chaos is quenched, such that the rheology becomes periodic and eventually steady. Thus, we illustrate that external flows have a dramatic influence on the dynamics of autophoretic disks, and, more generally, active droplets.

\section{\label{sec:gov_eq}Model and governing equations}

Our model consists of a circular disk of radius $a^{*}$ that is immersed in an incompressible Newtonian solvent of viscosity $\eta^{*}$ at temperature $T^{*}$. A steady simple shear flow with shear rate $\dot{\gamma}^{*}$ is imposed on the fluid, whose flow obeys the Stokes equations. The disk isotropically emits solute particles from its surface at a rate $\mathcal{A}^{*}$, and the solute diffusivity is $D^{*}$. The uniform solute concentration far away from the disk is $C^{*}_{\infty}$, and the excess solute concentration with respect to this far-field value is denoted by $c^{*}=C^{*}-C^{*}_{\infty}$. The disk interacts with the solute particles through a short-ranged potential of characteristic length $b^{*}$, which is much smaller than the radius of the disk~\cite{Anderson1989}. These interactions set up a tangential phoretic slip velocity, whose magnitude is determined by a mobility parameter $\mathcal{M}^{*}=\pm k_BT^{*}b^{*2}/\eta^{*}$, where $k_B$ is Boltzmann's constant, and the concentration gradient on the surface of the disk. The sign of the mobility parameter is positive (negative) for repulsive (attractive) interactions~\cite{Michelin2014}. Following~\cite{Golestanian2007,Michelin2013,Kailasham2022}, the scales for length, time, fluid velocity, pressure, and concentration are defined as $a^{*}$, ${a^{*}D^{*}}/{|\mathcal{A}^{*}\mathcal{M}^{*}|}$, $U^*={|\mathcal{A}^{*}\mathcal{M}^{*}|}/{D^{*}}$, ${\eta^{*}U^*}/{a^{*}}$, and ${a^{*}|\mathcal{A}^{*}|}/{D^{*}}$, respectively. It is convenient to define the scaled emission rate and mobility parameter as $A=\mathcal{A}^{*}/|\mathcal{A}^{*}|$ and $M=\mathcal{M}^{*}/|\mathcal{M}^{*}|$, respectively. The onset of phoretic self-propulsion requires {$AM>0$}~\cite{Michelin2013,Saha2021}, and we set $A=M=1$ henceforth. The intrinsic P\'{e}clet number quantifies the chemical activity of the disk and is defined as $Pe={a^{*}|\mathcal{A}^{*}\mathcal{M}^{*}|}/{D^{*2}}$. The dimensionless shear rate is given by $\epsilon=\dot{\gamma}^{*}{a^{*}D^{*}}/{|\mathcal{A}^{*}\mathcal{M}^{*}|}$, which can be viewed as a ratio of phoretic to flow time scales.

The fluid flow and solute concentration fields around the autophoretic disk placed in an ambient shear flow are calculated by simultaneously solving the transient advection-diffusion equation for the solute concentration, and the quasi-steady Stokes equations
\begin{equation}\label{eq:gov_eq}
Pe\left(\dfrac{\partial c}{\partial t} + \bm{v}\cdot\boldsymbol{\nabla}c\right) = \nabla^2 c; \boldsymbol{\nabla}p=\nabla^2\bm{v},\,\boldsymbol{\nabla}\cdot\bm{v}=0,
\end{equation}
with $t$, $\bm{v}$ and $p$ denoting the dimensionless time, velocity, and pressure respectively, and subject to $\left(\partial c/\partial r\right)(1,\theta,t)=-A$, and the attenuation condition $c(r\to\infty,t)\to0$. {The neglect of the time-derivative of the velocity field in~(\ref{eq:gov_eq}) is justified provided that the ratio of the kinematic viscosity of the fluid to the solute diffusivity is large~\cite{FNote1}, which is typically true for active droplet systems~\cite{Izri2014,Suda2021,Hokmabad2021,Chen2021}.} In numerical computations we employ a finite size of the computational domain, $R_{o}$, and impose the far-field Dirichlet condition $c(r=R_{o})=0$. Considering a frame-of-reference attached to the centroid of the disk, the slip velocity at the surface of the disk is given by $\bm{v}\left(r=1,\theta\right)=\bm{v}_{\text{s}}=M\nabla_{\text{s}}c$, with the surface-gradient operator defined as $\nabla_{\text{s}}=\left(1/r\right)\,\bm{e}_{\theta}\partial/\partial \theta$, where $\bm{e}_{\theta}$ is the unit vector in the tangential direction, $\theta$. The far-field velocity is given by $\bm{v}\left(R_{o},\theta\right)=-\bm{U}+\epsilon y\bm{e}_{x}$, where $\bm{e}_{x}$ the Cartesian unit vector in the $x$-direction, $y$ is the co-ordinate in the direction of the shear gradient, and the phoretic velocity $\bm{U}(t)$ may be evaluated from the slip velocity using the reciprocal theorem~\cite{Stone1996,Squires2006} as
\begin{equation}\label{eq:phor_vel}
\bm{U}(t)=-\dfrac{1}{2\pi}\int_{0}^{2\pi}\bm{v}_{\text{s}}d\theta.
\end{equation}
{The external shear acts to distort the solute field by contributing to the velocity field $\bm{v}$, which subsequently also affects the phoretic velocity $\bm{U}(t)$.}
The time-dependent base state, $c_{0}(r,\theta,t)$, corresponding to the state of zero phoretic motion ($\bm{U}(t)=0$) is computed by solving the transient advection-diffusion equation subject to an initial condition of $c_{0}(r,\theta,t=0)=0$ over a duration $t_{\text{b}}$, at specified $Pe$ and $\epsilon$. The value of the base state at $t_{\text{b}}$ is perturbed, $c(r=1,\theta,t=0)=c_{0}(r=1,\theta,t_{\text{b}})-\delta_{\text{per}}\cos\theta$, with $|\delta_{\text{per}}|<1$, and used as the initial condition while solving eq.~(\ref{eq:gov_eq}). A spectral element solver, described in detail in ~\cite{Chisholm2016,Kailasham2022}, is used for the numerical solution of~(\ref{eq:gov_eq}). 
\begin{figure}[t]
\begin{center}
\includegraphics[width=3.4in,height=!]{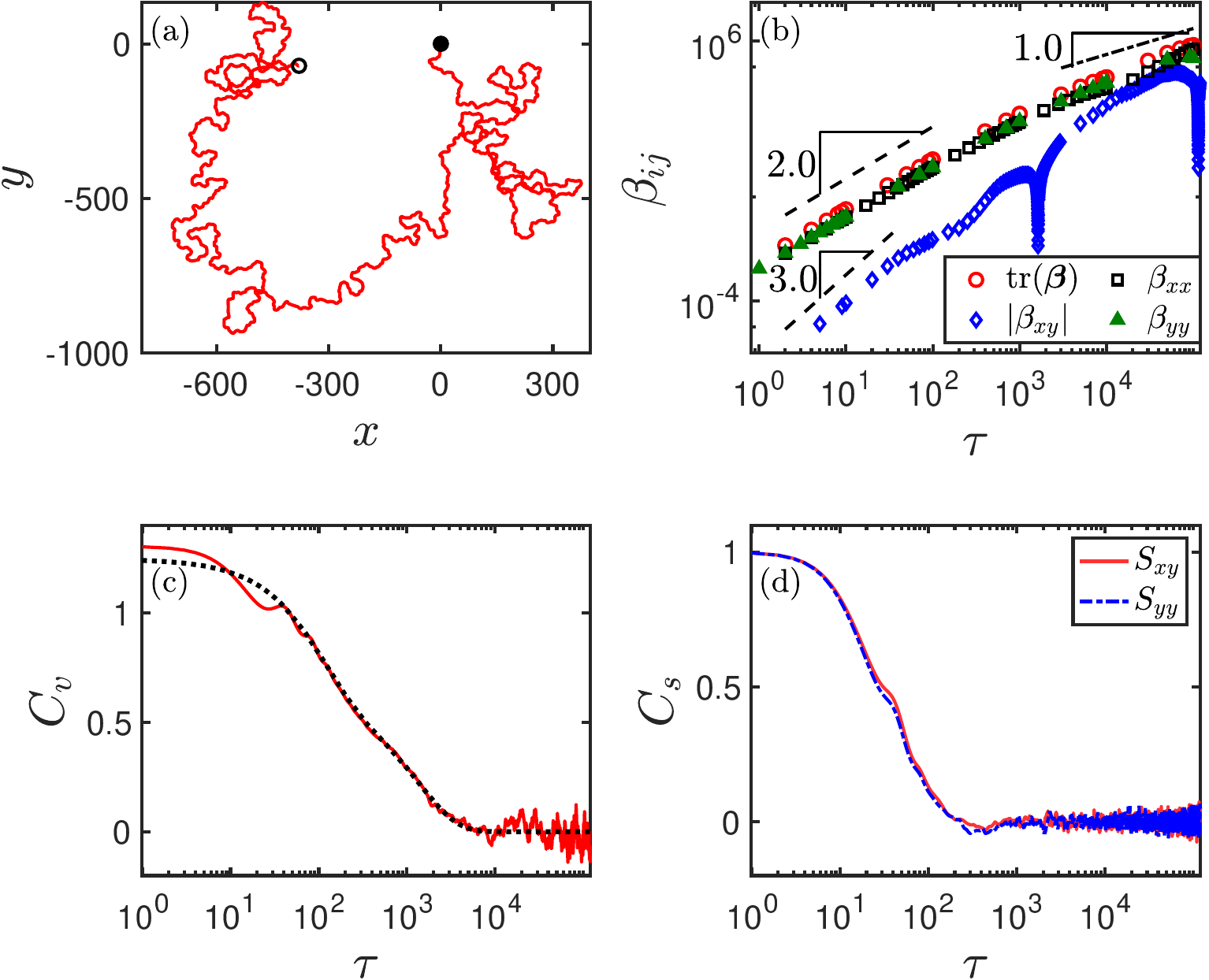}
\end{center}
\caption{(Colour online) Dynamics of a two-dimensional autophoretic disk in a quiescent fluid ($\epsilon=0$) at $Pe=20$. (a) Trajectory in the $x-y$ plane; solid and hollow circles denote the beginning and end of the trajectory, respectively, (b) elements of the mean square displacement tensor $\boldsymbol{\beta}(\tau)$, (c) velocity autocorrelation, and (d) stresslet component autocorrelation. The black dotted line in (c) has the functional form $f_1\exp[-\lambda_1\tau]+f_2\exp[-\lambda_2\tau]$, with the parameter values given by $f_1=0.64$, $\lambda_1=-0.009$, $f_2=0.60$, and $\lambda_2=-0.0007$.}
\label{fig:pe20_quie}
\end{figure}
\begin{figure}[t]
\begin{center}
\includegraphics[width=3.4in,height=!]{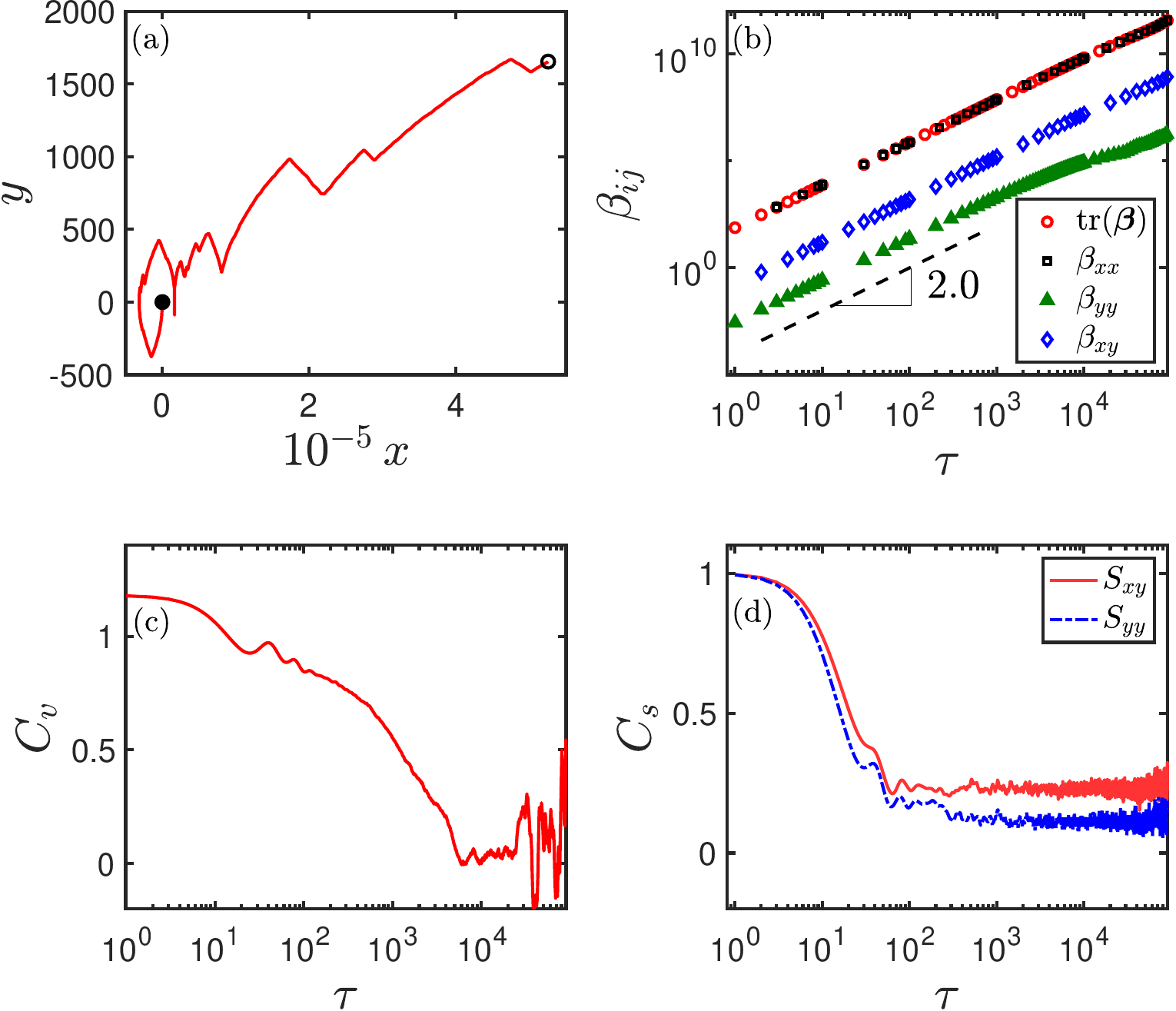}
\end{center}
\caption{(Colour online) {Dynamics of an autophoretic particle at $Pe=20$, placed in a shear flow field of strength $\epsilon=0.01$. The particle is initially at the origin, and the flow is in the $x$-direction only, with the velocity increasing linearly with $y$.} (a) Trajectory in the $x-y$ plane; (b) elements of the mean square displacement tensor $\boldsymbol{\beta}(\tau)$, (c) velocity autocorrelation, and (d) stresslet component autocorrelation.}
\label{fig:anltc_w_shear}
\end{figure}
The solution methodology for an active disk suspended in a quiescent fluid ($\epsilon=0$) is slightly different~\cite{FNote1}. A value of $R_{o}=200$ has been used in all the simulations, after establishing convergence~\cite{FNote1}. Using a procedure analogous to that in~\cite{Lauga2016}, the stresslet induced by the two-dimensional disk is derived as 
\begin{equation}\label{eq:stresslet_formula}
\bm{S}=-2\int_{0}^{2\pi}\left[\bm{n}\bm{v}_{\text{s}}+\bm{v}_{\text{s}}\bm{n}\right]d\theta,
\end{equation}
where $\bm{n}=\bm{e}_{r}$ is the unit outward normal on the surface of the disk. The expressions for the various components of the $2\times 2$ stresslet tensor have been derived in~\cite{FNote1}. Since $S_{xx}$ and $S_{yy}$ are identical in magnitude and differ only by sign, we only present results for $S_{xy}$ and $S_{yy}$ in this paper. In a dilute suspension of non-interacting autophoretic disks, $S_{xy}$ would contribute to the effective shear viscosity of the suspension.

\section{\label{sec:results}Results}

Results for an autophoretic disk suspended in a quiescent fluid ($\epsilon=0$) are presented first (fig.~\ref{fig:pe20_quie}), to provide context for our findings on disks placed in shear flow. The base state for a two-dimensional disk has no fluid flow and time-dependent solute diffusion, which we evaluate by numerically solving the transient diffusion equation. Our approach is in contrast to~\cite{Hu2019,Li2022} who assume a steady base state, in which the purely diffusive concentration field is set to zero at a finite distance $R_{o}$. In~\cite{FNote1} we show that both approaches yield qualitatively similar dynamics in the chaotic regime; however, there are important differences regarding the onset of self-propulsion.   

An autophoretic disk in a quiescent fluid undergoes qualitative changes in its trajectory as the P\'{e}clet number is increased, evolving from a stationary state to one of steady linear motion, followed by a meandering regime that transitions via an intermittency scenario to chaotic motion at sufficiently large $Pe$~\cite{FNote1,Hu2019}. The chaotic dynamics is characterized by a transition to diffusive scaling of the mean square displacement of the particle. We choose a representative value of $Pe=20$ that is deep in the chaotic regime. The mean square displacement tensor of the disk at a lag time $\tau$ is $\boldsymbol{\beta}(\tau)=\left<\left[\bm{r}(t+\tau)-\bm{r}(t)\right]\left[\bm{r}(t+\tau)-\bm{r}(t)\right]\right>$, where $\bm{r}(t)$ denotes the instantaneous position of the particle at time $t$, and the angular brackets represent the average evaluated over the trajectory of the particle. The off-diagonal components of this $2\times2$ tensor are identical, and the scalar mean square displacement is given by the trace of $\boldsymbol{\beta}$, i.e, $\text{MSD}\equiv\text{tr}(\bm{\beta})$. 

{It is seen from fig.~\ref{fig:pe20_quie}~(b) that} even though $\text{MSD}\sim\tau$ at long times, seemingly indicative of Brownian motion, the displacement cross correlation $\beta_{xy}$ does not vanish, which is markedly different from a passive Brownian particle. To represent the data on logarithmic axes, we plot the absolute values of the off-diagonal term, $\beta_{xy}$. The growth of $\beta_{xy}$ with respect to the lag time proceeds with abrupt dips, which have been observed by~\citet{Suga2018} in their experiments on active liquid crystal droplets swimming in a surfactant solution in a two-dimensional geometry. {While the overall MSD for the liquid crystal droplets oscillates in time it is the off-diagonal component of the MSD tensor which displays an oscillatory behavior in the present work. ~\citet{Suga2018} observe that the oscillation in the MSD corresponds to a case where the droplet traces out multiple ``figure-8" or loop-like patterns. An examination of the disk trajectory in our work (fig.~\ref{fig:pe20_quie}~(a)), however, reveals that such loop-like patterns constitute only a small fraction of the overall trajectory. {There is an early-stage $\sim\tau^3$ scaling observed for $\beta_{xy}$ whose origins remain unclear.} The overall MSD of the autophoretic disk in a quiescent fluid is therefore not oscillatory, but rather exhibits the ballistic-diffusive transition observed previously in numerical investigations~\cite{Hu2019,Hu2022}. The off-diagonal component of the MSD tensor for self-propelled objects has not been examined widely in the literature. ~\citet{TenHagen2011} consider an active Brownian particle (ABP) model in which the orientation of the particle has a deterministic component stemming from a finite angular velocity, as well as a stochastic component. They present a detailed derivation for the mean square displacement of the particle. Following a similar route, it may be shown that $\beta_{xy}$ is zero in the absence of an angular velocity, but does not vanish in general for finite values of the angular velocity. This agrees with the intuition that the $x$- and $y$-displacements of a particle moving in two-dimensions would be correlated in presence of a finite, deterministic angular velocity.} The velocity and stress autocorrelations are evaluated as $C_{v}(\tau)=\left<\bm{U}(t)\cdot\bm{U}(t+\tau)\right>/\left<|\bm{U}(t)|\right>^2$, and $C_{s}(\tau)=\left<s(t)s(t+\tau)\right>/\left<s^2(t)\right>$, respectively, where $s$ represents either $S_{xy}$ or $S_{yy}$. We observe that $C_{v}$ in fig.~\ref{fig:pe20_quie}~(c) is well-approximated by a sum of two exponentials. The autocorrelation of both the $xy$ and $yy$ components of the stresslet decay nearly identically, going to zero at $\tau\approx 300$ (fig.~\ref{fig:pe20_quie}~(d)). This is to be expected because there is no ambient flow that would lead to a distinction between the components of the stresslet.

\begin{figure}[t]
\centering
\includegraphics[width=3.4in,height=!]{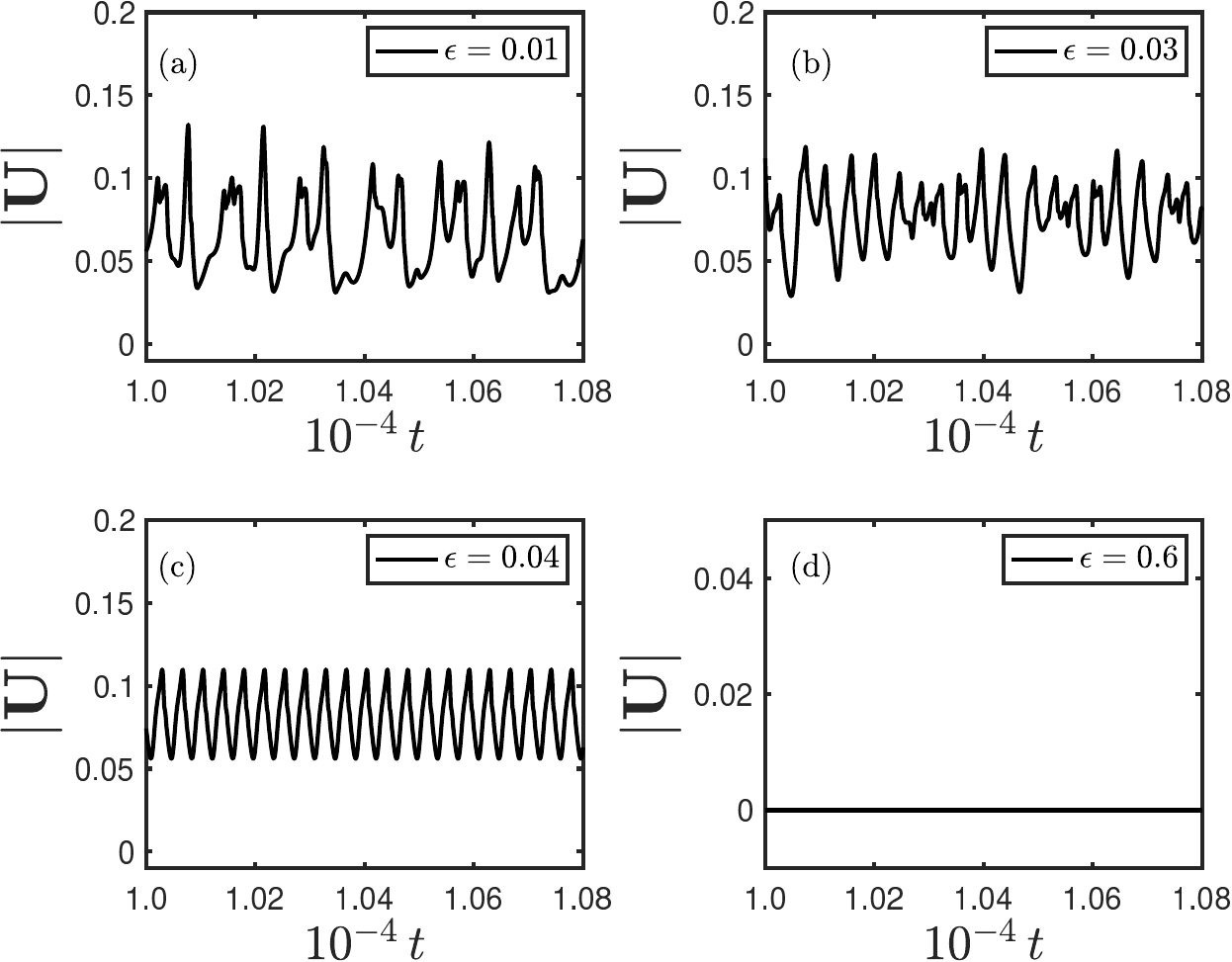}
\caption{Time series of velocity magnitude at shear rate values of (a) $\epsilon=0.01$, (b) $\epsilon=0.03$, (c) $\epsilon=0.04$,  and (d) $\epsilon=0.6$, at a fixed value of the P\'{e}clet number, $Pe=20$.}
\label{fig:sr_effect_vel}
\end{figure}
\begin{figure}[t]
\begin{center}
\includegraphics[width=3.4in,height=!]{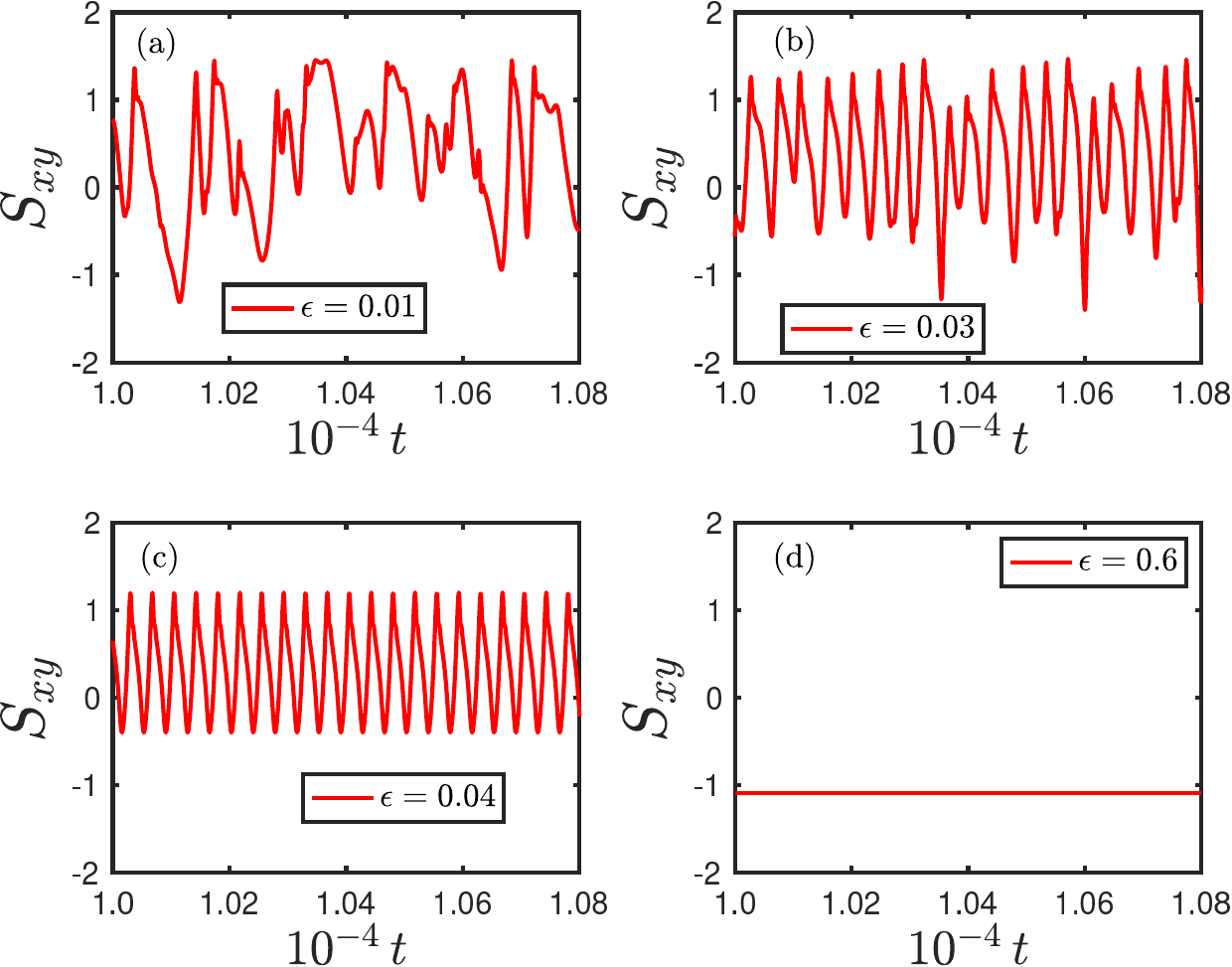}
\end{center}
\caption{(Colour online) Time series of $xy$ component of the stresslet at shear rate values of (a) $\epsilon=0.01$, (b) $\epsilon=0.03$, (c) $\epsilon=0.04$,  and (d) $\epsilon=0.6$, at a fixed value of the P\'{e}clet number, $Pe=20$.}
\label{fig:sr_effect_slet}
\end{figure}

A widely used reduced-order framework for self-propelled micro-scale objects is the ABP~\cite{Bechinger2016,Marchetti2016,Zeitz2017}, wherein a particle is assumed to move with a constant speed, while its instantaneous orientation is selected from a Gaussian white noise distribution. ~\citet{Peruani2007} (hereafter PM) prescribe a model for a self-propelling particle in two dimensions which allows for fluctuations in both the magnitude and direction of velocity. By assuming that the instantaneous velocity magnitude may be selected stochastically from a Poisson distribution, and the orientational distribution function obeys a diffusion equation, they analytically derive an expression for MSD and $C_v$ of such a particle. Both the ABP and PM models predict an MSD that transitions to long-time diffusion following an early-time ballistic behavior. However, while the ABP predicts that a single-exponential may be used to describe the variation in $C_v$, the PM model posits that the velocity autocorrelation is a sum of two exponentials. In fig.~\ref{fig:pe20_quie}~(c), we show agreement with this bi-exponential form~\cite{Peng2021} and the computed velocity autocorrelation of an autophoretic disk. {It is therefore evident that the velocity time-series of an autophoretic disk undergoing chaotic dynamics is better described by the PM model that allows for the instantaneous velocity and orientation to be chosen independently, rather than the ABP model.}

For a passive Brownian sphere in shear flow, the mean square displacement in the flow direction ($\beta_{xx}$) scales as $\tau^3$ at long times, whereas that in the direction of the {shear} gradient ($\beta_{yy}$) scales linearly with the lag time~\cite{Sandoval2014,Foister1980,Rhines1983}. The cross-correlation, $\beta_{xy}$, scales as $\tau^2$. The effect of a shear-flow field on the dynamics of an autophoretic disk is illustrated in fig.~\ref{fig:anltc_w_shear}. The position of the autophoretic disk is evaluated as, 
\begin{equation}
\begin{split}
r_{y}(t+\Delta t)&=r_{y}(t)+U_{y}(t)\Delta t;\\
r_{x}(t+\Delta t)&=r_{x}(t)+\left[U_x(t)+\epsilon r_{y}(t)\right]\Delta t,
\end{split}
\end{equation}
while the expression for the mean square displacement tensor remains unchanged. In fig.~\ref{fig:anltc_w_shear}~(b), the components of $\boldsymbol{\beta}(\tau)$ for an autophoretic disk placed in a shear-flow field of $\epsilon=0.01$ are plotted as a function of the lag time, and it is observed that they all scale as $\tau^2$, in marked contrast to the scaling observed for a passive disk. The autocorrelation of the phoretic velocity is not significantly altered in comparison to the quiescent case (fig.~\ref{fig:anltc_w_shear}~(c)). The autocorrelation of the stresslet components (fig.~\ref{fig:anltc_w_shear}~(d)) do not vanish in the long-time limit due to the presence of an ambient shear flow. Furthermore, the $xy$ and $yy$ components of the stress tensor may be clearly differentiated from their autocorrelation signals, unlike that in the quiescent case, since the imposed flow has a fixed direction, along the $x$-axis.

\begin{figure}[t]
\begin{center}
\includegraphics[width=3.in,height=!]{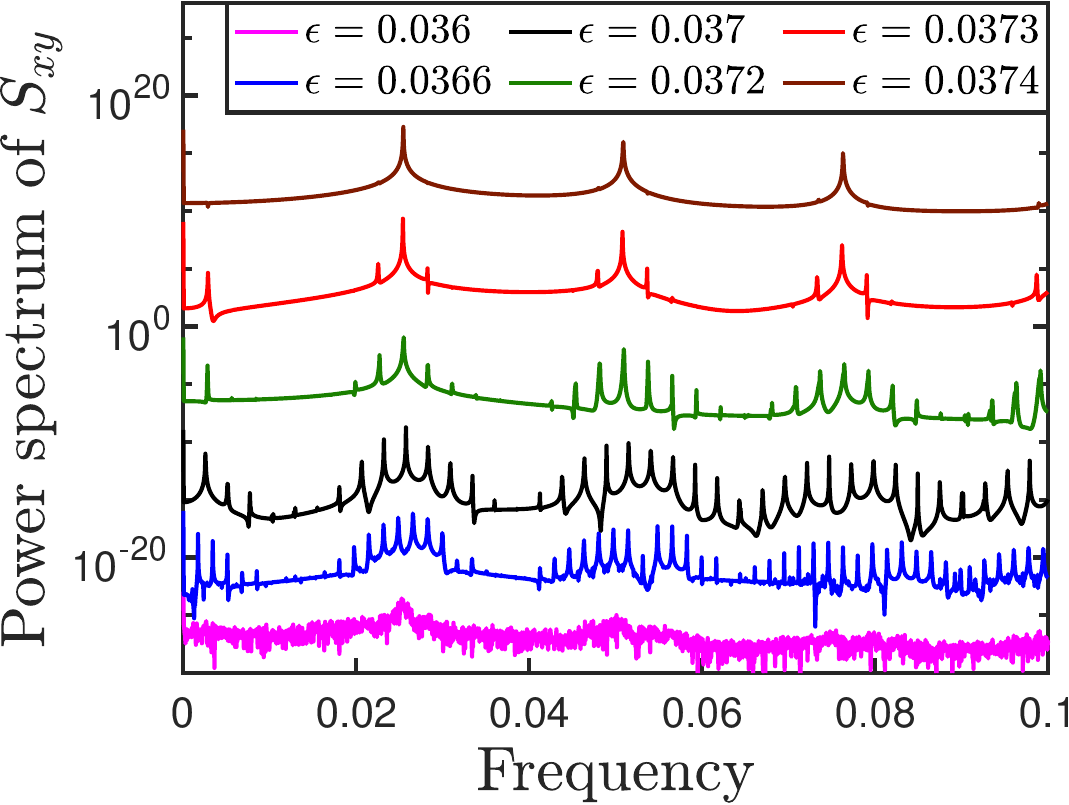}
\end{center}
\caption{(Colour online) Power spectrum of the $xy$ component of the stresslet at various values of the shear rate ($\epsilon$), and a fixed P\'{e}clet number of $Pe=20$. The ordinates of the data-series have been multiplied by a scale factor to render them well-spaced on the $y$-axis, for clarity. From bottom to top, the values of $\epsilon$ (and the associated scale factors) are 0.036 ($10^{-22}$), 0.0366($10^{-15}$), 0.037($10^{-8}$), 0.0372(1), 0.0373($10^{10}$), and 0.0374($10^{18}$).}
\label{fig:pspec_sxy}
\end{figure}

{A strategy for quenching the chaotic motion of the autophoretic disk is discussed next. In the context of high-Reynolds number applications, the use of magnetic fields~\cite{Moffatt1967}, buoyant forces~\cite{He2016,Marensi2021}, and mechanical impulses~\cite{Kuhnen2018} has been suggested to relaminarize the flow~\cite{Hanratty2009}.} {In a similar spirit, recent simulations~\cite{Kailasham2022} and experiments~\cite{Castonguay2023} have established that an external force field may be used to quench the chaotic variation of the velocity field around an autophoretic particle~\cite{Stark2016}.} In fig.~\ref{fig:sr_effect_vel}, the effect of increasing the shear strength on the phoretic velocity of the particle at a fixed value of the P\'{e}clet number is illustrated, over a representative time window. At low values of $\epsilon$, the time-series is chaotic, and then gradually settles into a periodic pattern upon increasing $\epsilon$, before vanishing completely as the shear rate is increased further. The triangular wave pattern of the phoretic velocity in fig.~\ref{fig:sr_effect_vel}~(c) is qualitatively similar to that predicted for self-propelled two-dimensional droplets~\cite{Li2022}. The external flow field, therefore, modulates the self-propulsive motion of the autophoretic disk. A similar trend is observed in fig.~\ref{fig:sr_effect_slet} for the stresslet component $S_{xy}$, which notably attains a time-independent value at sufficiently large shear rates. A negative value for the steady-state stresslet has also been observed for the case of axisymmetric swimmers with a ``pusher'' type flow pattern~\cite{Michelin2013,Lauga2016}. We therefore anticipate that a dilute suspension of active droplets would exhibit a chaotic rheology at small values of the shear rate, before first transitioning to a time periodic and ultimately steady rheology as the shear rate is increased.

A more detailed picture of the chaotic-to-steady quenching transition is apparent from the power spectrum of the stresslet, as plotted in fig.~\ref{fig:pspec_sxy} over a narrow window in the shear rate. At $\epsilon=0.036$, the broadband spectrum signifies chaotic dynamics~\cite{Ikeda1980,Lai1998,hilborn2000}. An increase in the shear rate is accompanied by a coalescence of various frequencies from the broadband spectrum into discrete peaks of the periodic motion in fig.~\ref{fig:sr_effect_slet}~(c). {The regularly spaced spikes in the power spectrum at higher values of the dimensionless shear rate, however, do not coincide with integral multiples of $\epsilon$~\cite{FNote1}.}

\section{\label{sec:concl}Conclusions}

We have shown that the dynamics of an autophoretic disk is markedly different from a passive Brownian particle. The dynamics is also richer than that of an ABP with respect to the bi-exponential decay of the velocity autocorrelation function. Furthermore, we showed that the rheology of a dilute suspension of such disks is chaotic at small values of the shear rate; effectively, increasing the shear provides a route to removing chaotic dynamics. We expect these findings to hold, at least qualitatively, in experimentally realizable {dilute} active droplet systems, upto an area fraction of $\phi\lesssim 10\%$ based on prior studies on hard-sphere colloidal dispersions~\cite{Foss2000}. Finally, our work may be useful for the synthesis and manipulation of active droplet emulsions. Here, one may have to account for the coupled chaotic dynamics of many active droplets, which is an interesting direction for future work.
\begin{acknowledgments}
We gratefully acknowledge the support of the Charles E. Kaufmann Foundation of the Pittsburgh Foundation (Grant No. 1031373-438639).
\end{acknowledgments}

\bibliography{ms}
\end{document}


\beginsupplement

\newcommand{\blueline}{\raisebox{2pt}{\tikz{\draw[-,blue,solid,line width = 1.5pt](0,0) -- (5mm,0);}}}

\newcommand{\blackline}{\raisebox{2pt}{\tikz{\draw[-,black,dashed,line width = 1.5pt](0,0) -- (5mm,0);}}}



\title{Supplementary Material for: Non-Brownian diffusion and chaotic rheology of autophoretic disks}


\author{R. Kailasham}
\affiliation{Department of Chemical Engineering, Carnegie Mellon University, Pittsburgh, PA 15213, USA}
\author{Aditya S. Khair}
\email{akhair@andrew.cmu.edu}
\affiliation{Department of Chemical Engineering, Carnegie Mellon University, Pittsburgh, PA 15213, USA}


\date{\today}


\maketitle


\section{\label{sec:intro}Introduction}

The main paper solves the Stokes equation and transient advection-diffusion equation for the velocity and concentration fields around an autophoretic disk, and makes predictions about the dynamics of the particle as well as the rheology of a dilute suspension of non-interacting disks. Additional details pertaining to the study are presented here.

This document is organized as follows. A justification for the use of an autophoretic disk as a model for active droplets moving in three dimensions is presented in Section~\ref{sec:model_justif}. The special case of $\epsilon=0$, corresponding to an active two-dimensional disk in a quiescent fluid is discussed in Section~\ref{sec:quie_solve}. Here, we discuss how our solution method for this case differs from~\cite{Hu2019}. While we do not find a significant qualitative difference in the computational predictions within the chaotic regime, the difference in approaches is worth discussing. The convergence of simulation results with respect to the size of the computational domain and the timestep width is discussed in Section~\ref{sec:num_conv}. Expressions for the components of the stresslet $\bm{S}$ are provided in Section~\ref{sec:slet_exp}, while Section~\ref{sec:power_spec} provides additional 
details on the power spectrum of the $xy$-component of the stresslet, for values of the dimensionless shear rate ($\epsilon$) close to the transition from chaotic to steady rheology.


\section{\label{sec:model_justif}Autophoretic disk as a model for an active droplet}

In this work, we consider a two-dimensional autophoretic disk as a minimal model for an active drop moving in three dimensions. The transition in the trajectory of an active droplet from steady self-propulsion at small $Pe$ to increasingly convoluted patterns at sufficiently large $Pe$ is a common feature that has been observed experimentally in two and three dimensions~\cite{Suga2018}. This trend has also been predicted qualitatively by simulations of autophoretic disks~\cite{Hu2019} and droplets~\cite{Li2022} in two dimensions, as well as rigid spheres in three dimensions~\cite{Hu2022}. The appearance of a long-time diffusive regime in the MSD of an autophoretic particle, following an early-time ballistic behavior, has also been observed in simulations of both rigid disks~\cite{Hu2019} and spheres~\cite{Hu2022}. Through numerical simulations of an active drop moving in unsteady rectilinear translation, ~\citet{Morozov2019} examined the relative magnitudes of diffusiophoretic and Marangoni effects on the transition to chaos, and suggest that this transition might not occur for purely Marangoni flows; a small amount of diffusiophoresis is needed to precipitate the transition. Numerical simulations of rigid autophoretic colloids indicate that increasing the strength an external biasing force quenches the transition to chaos~\cite{Kailasham2022}, a finding that has since been corroborated through experiments of active droplets settling under gravity~\cite{Castonguay2023}. We therefore believe that the use of autophoretic disks to model active droplets is justified for qualitatively predicting not only the changes in trajectory as $Pe$ is varied, but also the influence of an external agent (in our case a shear field) on modulating the transition to chaos. In particular, a core prediction of the present work, that the chaotic dynamics of an active droplet in a shear flow field is quenched at high values of the shear rate, should hold for other systems (e.g., self-diffusiophoretic Janus particles~\cite{Michelin2014}) whose mechanism for self-propulsion differs from that considered in the present work.

\section{\label{sec:quie_solve}Autophoretic disk in a quiescent fluid}

The dimensionless governing equations for the concentration and flow fields around an autophoretic disk are as follow,
\begin{equation}\label{eq:gov_eq}
Pe\left(\dfrac{\partial c}{\partial t} + \bm{v}\cdot\boldsymbol{\nabla}c\right) = \nabla^2 c; \boldsymbol{\nabla}p=\nabla^2\bm{v},\,\boldsymbol{\nabla}\cdot\bm{v}=0,
\end{equation}
subject to $\left(\partial c/\partial r\right)(1,\theta,t)=-A$ and $c(r\to\infty)=0$. For numerical purposes, the computational domain is truncated at a radius $R_{o}$, and the attenuation criterion is imposed as a Dirichlet condition in the far-field, i.e., $c(r=R_{o})=0$. A value of $R_{o}=200$ has been chosen in all our simulations, after testing for convergence of results as discussed in Section~\ref{sec:num_conv}. The velocity boundary conditions are $\bm{v}\left(r=1,\theta\right)=\bm{v}_{\text{s}}=M\nabla_{\text{s}}c$ and $\bm{v}\left(R_{o},\theta\right)=-\bm{U}$, where the surface-gradient operator is defined as $\nabla_{\text{s}}=\left(1/r\right)\,\bm{e}_{\theta}\partial/\partial \theta$, with $\bm{e}_{\theta}$ denoting the unit vector in the tangential direction.  An expression for the phoretic velocity, $\bm{U}(t)$, may be written, using the reciprocal theorem as follows~\cite{Stone1996,Squires2006},
\begin{equation}
\bm{U}(t)=-\dfrac{1}{2\pi}\int_{0}^{2\pi}\bm{v}_{\text{s}}d\theta.
\end{equation}
Neglecting the time-derivative $(\rho\,\partial_t v)$ in the Stokes equations is valid provided that the ratio of the kinematic viscosity of the fluid $\nu$ to the solute diffusivity $D$ (quantified as the Schmidt number, $Sc$) is large. An analysis of the experiments by~\citet{Hokmabad2021} on self-propelled oil droplets in an aqueous surfactant yields values of $\nu\approx10^{-5}\text{m}^2/\text{s}$ and $D\approx8.7\times10^{-12}\text{m}^2/\text{s}$, giving $Sc\approx\,10^{6}$. For water droplets moving in a monoolein-squalane mixture~\cite{Izri2014,Suda2021}, we find that $\nu\approx3.8\times10^{-5}\text{m}^2/\text{s}$ and $D\approx2.3\times10^{-9}\text{m}^2/\text{s}$~\cite{Suda2021}, giving $Sc\approx\,10^{4}$. It is therefore clear that the fluid flow around these active droplets may be modeled by the quasi-steady Stokes equations that does not contain the time derivative of the velocity field. Theoretical and computational works on this topic~\cite{Michelin2013,Morozov2019,Kailasham2022,Michelin2023} have routinely followed this practice, and successfully predict the transition to chaos in the trajectory of the self-propelled droplets, by setting $\left(1/Sc\right)=0$. Finally, we note that~\citet{Chen2021} use the full Navier-Stokes equations and investigates the effect of a finite Schmidt number on the dynamics of an autophoretic colloid moving in three dimensions. In Fig.~12 of their work, it is seen that their predictions for the phoretic velocity over a range of P\'{e}clet numbers matches with the $Sc\to\infty$ results obtained by~\citet{Michelin2013}, who use the Stokes equations and assume that the colloid moves in an axisymmetric geometry. In the same figure, ~\citet{Chen2021} observe that the prediction for the phoretic velocity remains largely invariant as the Schmidt number is increased beyond $Sc\geq 1$, for $Pe=8,12$. They have used $Sc=1$ in all their simulations of an autophoretic colloid. Furthermore,~\citet{Chen2021} observe a transition to chaotic dynamics at high enough values of $Pe$. It is therefore clear that despite the neglect of inertial effects, the essential features of the self-propelled motion of an active droplet are recovered in the current model.

The self-propelled motion of the disk arises due to the phoretic interaction between the surface of the disk and the solute particles emitted from it. When the range of the interaction potential is small compared to the size of the disk, it is customary to assume a slip boundary condition for the velocity field at the surface of the disk~\cite{Anderson1989,Golestanian2007,Chen2021,Saha2021,Michelin2023}. This approximation is also made in models for self-propelled droplets that successfully capture the features observed in experiments, as seen from refs.~\cite{Izri2014,Hokmabad2021}. An exact numerical value for the range of the interaction potential would depend on the specificity of the interaction (e.g. Coulombic) but a reasonable estimate would be given by the size of the emitted solute particles, or swollen micelles for the case of active droplets. For example, in the case of CB15 droplets moving in an aqueous solution of the surfactant TTAB~\cite{Hokmabad2021}, the contrast between the radius of the droplet ($\sim30\,\mu\text{m}$) and that of the filled micelle ($\sim2.5\,\text{nm}$) is a factor of approximately $10^{-4}$. This justifies the use of the slip boundary condition approximation even for droplet systems. 

For a given value of $Pe$, the first step in our solution methodology involves the numerical computation of the unsteady isotropic base state $c_{0}(r,t)$ corresponding to no phoretic motion, i.e., $\bm{U}(t)=0$. We compute the evolving solute concentration in this base state by solving the initial value problem given by
\begin{equation}\label{eq:trans_diff}
Pe\left(\dfrac{\partial c_{0}}{\partial t}\right)=\nabla^2 c_{0}
\end{equation}
subject to $\left(\partial c_0/\partial r\right)(1,t)=-A$, $c_0(r\to\infty)=0$, and $c_0(r,t=0)=0$. A uniformly valid asymptotic approximation for the concentration field may be obtained in the low $Pe$ limit using the method of matched asymptotic expansion. The detailed derivation is available upon request from the authors, and we only state the final result for the concentration field below,
\begin{equation}\label{eq:us_base_state}
c_0(r,t)\sim\dfrac{A}{2}\int_{\rho^2/4t}^{\infty}u^{-1}e^{-u}du;\quad \rho=Pe^{1/2}r.
\end{equation}
\begin{figure}[h]
\begin{center}
\begin{tabular}{c c}
\includegraphics[width=3in,height=!]{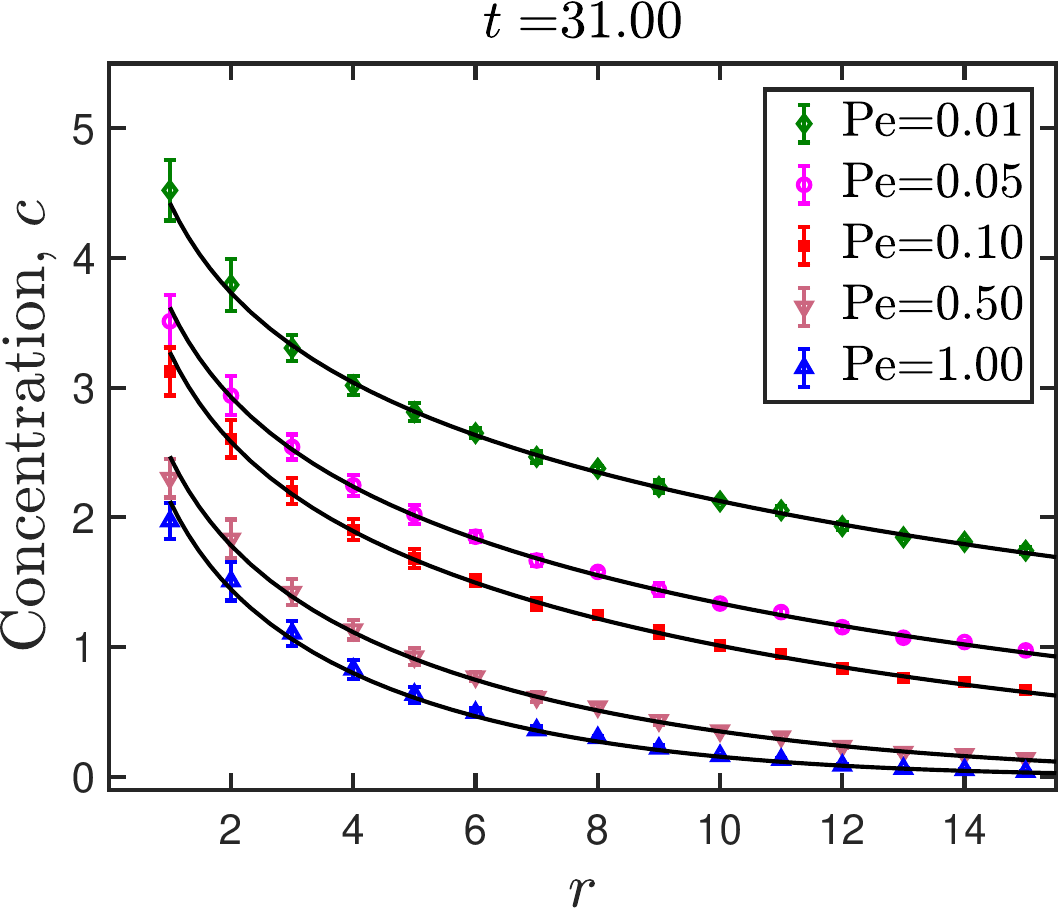}&
\includegraphics[width=3in,height=!]{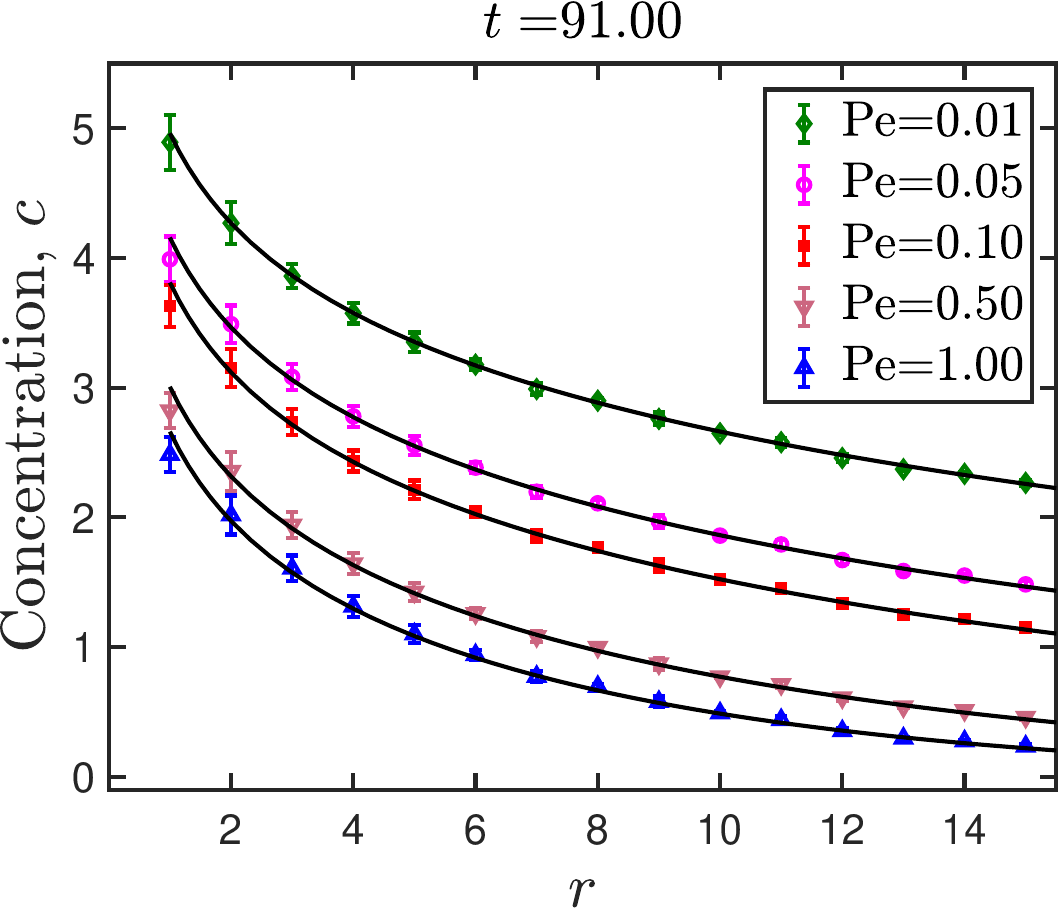}\\
(a)&(b)
\end{tabular}
\end{center}
\caption{(Colour online) Radial concentration profile at several values of the P\'{e}clet number, at (a) $t=31$, and (b) $t=91$. Symbols are numerical results, while lines represent eq.~(\ref{eq:us_base_state}). Coordinates in the $r$-direction are discretized into bins of width 1 unit. The concentration values falling within this bin are
averaged over and plotted against the midpoint of the bin location. Error bars on each data point indicate the standard deviation of concentration values within that bin.}
\label{fig:compare_us_base_state}
\end{figure}
In Fig.~\ref{fig:compare_us_base_state}, the results obtained by computing the integral in eq.~(\ref{eq:us_base_state}), for several small values of the P\'{e}clet number, is compared against the numerical solution of eq.~(\ref{eq:trans_diff}) at two different values of time. The excellent agreement between the results establishes the validity of the asymptotic approximation and the robustness of our numerical simulations.

As the next step, the total concentration field is expressed as the sum of the time-dependent base state and a fluctuation, 
\begin{equation}\label{eq:split_c}
c(r,\theta,t)=c_{0}(r,t)+c'(r,\theta,t)
\end{equation}
Substituting eq.~(\ref{eq:split_c}) into the transient advection-diffusion eq.~(\ref{eq:gov_eq}) and simplifying using eq.~(\ref{eq:trans_diff}), the governing equation for $c'$ is obtained as follows
\begin{equation}
Pe\left(\bm{v}\cdot\boldsymbol{\nabla}c_{0}\right)+Pe\left[\dfrac{\partial c'}{\partial t}+\bm{v}\cdot\boldsymbol{\nabla}c'\right]=\nabla^2 c'
\end{equation}
with a no-flux boundary condition for $c'$ at the surface of the disk, and a vanishing concentration at the edge of the computational domain. The boundary conditions for the velocity field are identical to those given below eq.~(\ref{eq:gov_eq}), with $c$ replaced by $c'$. The initial condition for the concentration fluctuation is given by a perturbation applied only to the surface of the particle, as follows
\begin{equation}
c'(r=1,\theta,t=0)=-\delta_{\text{per}}\cos\theta;\quad |\delta_{\text{per}}|<1
\end{equation}
We term this solution methodology, which involves numerically calculating the time-dependent base-state, as the ``transient base state" approach.

We next recap an $ad\,hoc$ methodology that has been used in the literature for autophoretic disks~\cite{Hu2019,Li2022}. We shall term this the ``stationary base state" approach for reasons that will be apparent. First, recall that there is no solution to the steady diffusion (i.e., Laplace's) equation in two dimensions that permits a vanishing far-field concentration as $r\to\infty$. In a finite-sized geometry, however, there exists a steady concentration profile, 
\begin{equation}\label{eq:stat_base}
c_{0}=A\ln\left(R_{o}/r\right)
\end{equation}
at all values of the P\'{e}clet number, corresponding to a base state of zero phoretic velocity, and no fluid flow. In the stationary base-state approach adopted in~\cite{Hu2019,Li2022}, a perturbation is applied to the base state given by eq.~(\ref{eq:stat_base}), and the transient advection-diffusion equation is solved. That is, in those studies, the lack of a solution to Laplace's equation in an unbounded domain is circumvented by an imposed cutoff at $R_{o}$. This enables perturbation about a steady base state. Properly, however, the base state for an autophoretic disk is one of transient diffusion, as we have adopted above. Perturbations about the steady base state (\ref{eq:stat_base}) are amplified for values of $Pe>Pe_{1}$, where the critical P\'{e}clet number associated with the transition from stationarity to linear self-propulsion depends on the outer radius of the computational domain as $Pe_{1}\sim2/\ln R_{o}$~\cite{Hu2019}.

\begin{figure}
\centering
\includegraphics[width=5in,height=!]{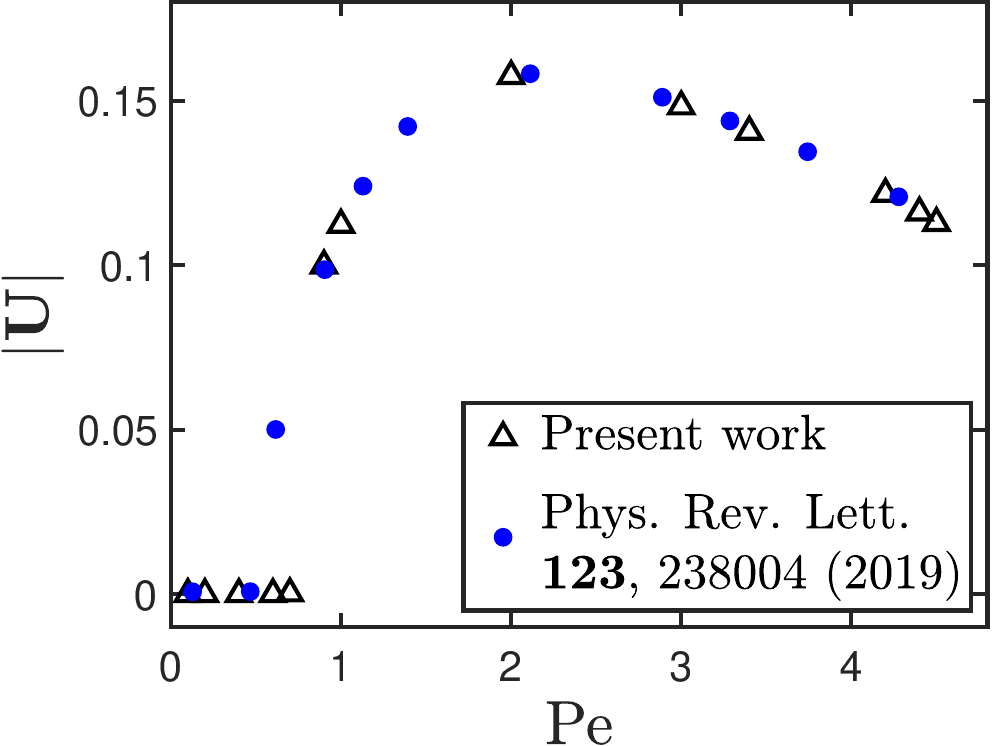}
\caption{Magnitude of the phoretic velocity of a 2D active particle in a quiescent fluid, obtained using the transient base state approach employed in the present work, and compared against the results of~\citet{Hu2019} that assumes a stationary base state. }
\label{fig:u_v_pe_validn}
\end{figure}

In fig.~\ref{fig:u_v_pe_validn}, the magnitude of the phoretic velocity obtained using our transient base state approach is compared against the values reported in~\citet{Hu2019} that employs the stationary base state assumption. The excellent agreement between the results indicates not only the validity of our computational method, but also suggests that the stationary base-state assumption is perhaps not a severe one. There is, however, a difference in the predictions of the two approaches for the value of $Pe_{1}$, or the first bifurcation point at which the transition from stationary state to steady self-propulsion is observed. 
\begin{figure}
\centering
\includegraphics[width=5in,height=!]{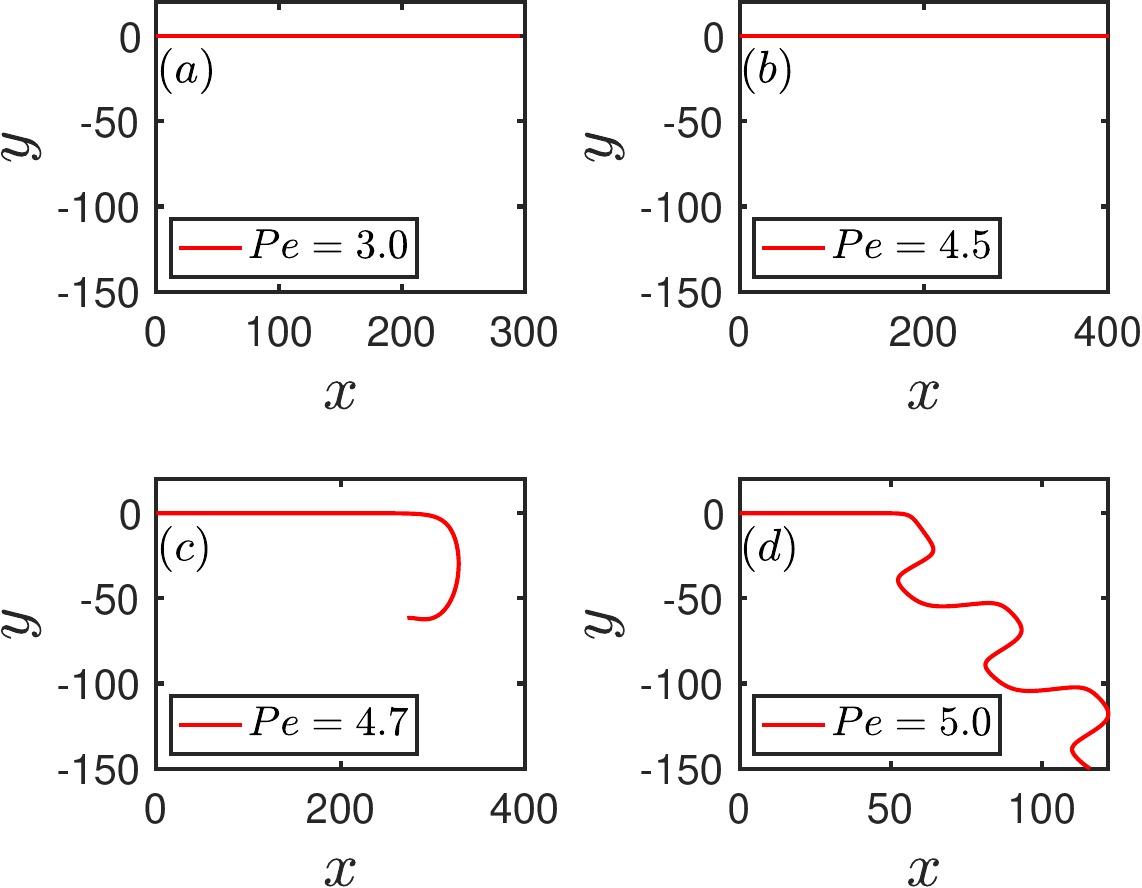}
\caption{Trajectory of a 2D autophoretic particle in a quiescent fluid, computed at (a) Pe=3, (b) Pe=4.5, (c) Pe=4.7, and (d) Pe=5.}
\label{fig:traj_plot_pe_2}
\end{figure}
While the stationary base state approach predicts $Pe_{1}\sim0.466$, results from the transient-base state approach predict that the particle does not move until $Pe>0.7$. We believe that this difference stems from our differing approaches for the solution of the governing equation: while we truncate the computational domain at $R_{o}$ for numerical reasons, we solve for both the velocity and the concentration fields simultaneously.~\citet{Hu2019} truncate only the concentration field and use the streamfunction solution~\cite{Blake1971,Sondak2016} that is strictly valid for an infinite domain. Despite these differing solution techniques, the two methods do agree on their predictions for $Pe_{2}\sim4.65$, the P\'{e}clet number at which the transition from linear to meandering motion occurs. From the plots of the $x-y$ trajectories in fig.~\ref{fig:traj_plot_pe_2}, it is evident that the transition from linear to meandering motion sets in at a value of $Pe$ between $4.5$ and $4.7$.

\begin{figure}[t]
\begin{center}
\includegraphics[width=3.5in,height=!]{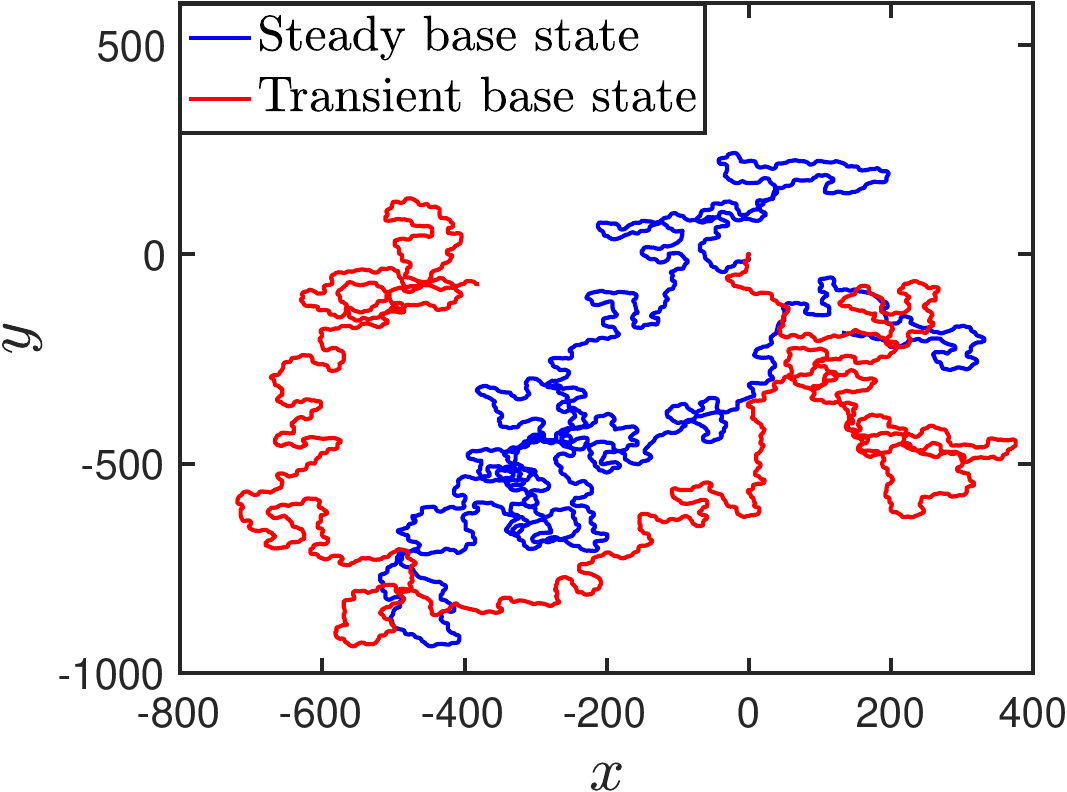}\\
(a)\\[15pt]
\begin{tabular}{cc}
\includegraphics[width=3in,height=!]{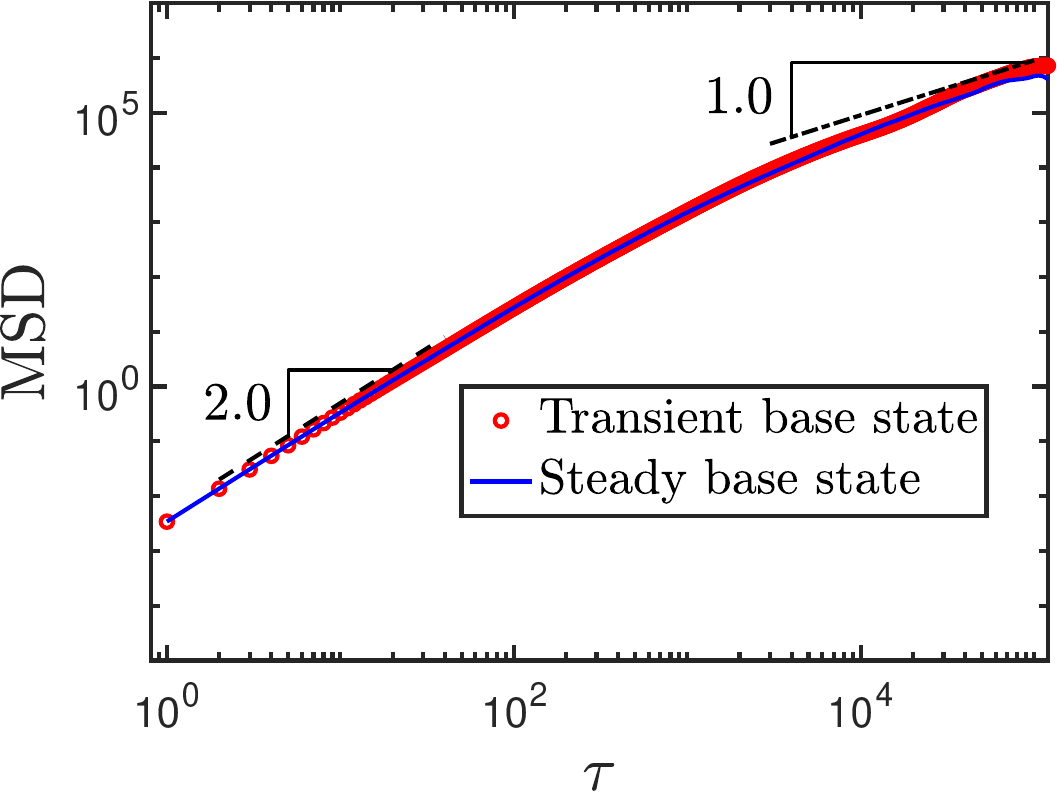}&
\includegraphics[width=3in,height=!]{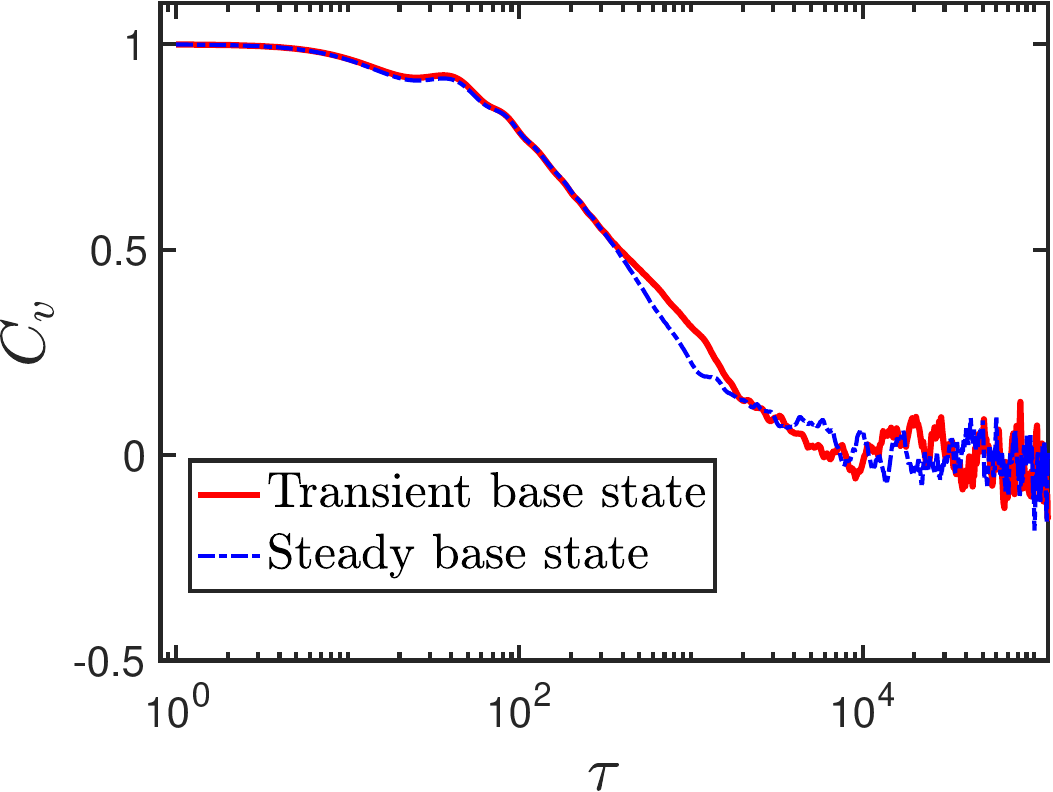}\\
(b)&(c)
\end{tabular}
\end{center}
\caption{(Colour online) (a) Trajectory in the $x-y$ plane, (b) mean square displacement, and (c) velocity autocorrelation for a two dimensional autophoretic particle in a quiescent fluid at Pe=20.}
\label{fig:pe20_compare_approach}
\end{figure}

As the P\'{e}clet number is increased further, the stationary base state approach~\cite{Hu2019} predicts that the dynamics becomes chaotic at $Pe\approx 13$, which is marked by a transition to diffusive scaling in the mean square displacement at long times. Using the transient base state approach, however, we do not observe a long-time-scale diffusion at $Pe=13$. We pick a value of $Pe=20$, deep in the chaotic regime, and simulate the particle dynamics using both the steady and transient base-state approaches. In both the approaches, we do not use the known streamfunction solution, but rather solve for both the velocity and concentration fields simultaneously. The resulting trajectory, mean-square displacement, and velocity autocorrelation from the two methods are compared in fig.~\ref{fig:pe20_compare_approach}. While the actual trajectory in the $x-y$ plane for the two cases are different, it is readily apparent that the MSD and the velocity autocorrelation are nearly identical. This further indicates that the approximation of a stationary base state works well for the case of a two dimensional autophoretic disk in a quiescent fluid. All results for the autophoretic disk in a quiescent fluid presented in the main text have been obtained using the transient base state approach.

\section{\label{sec:num_conv}Numerical convergence}

\begin{figure}[h]
\centering
\includegraphics[width=5in,height=!]{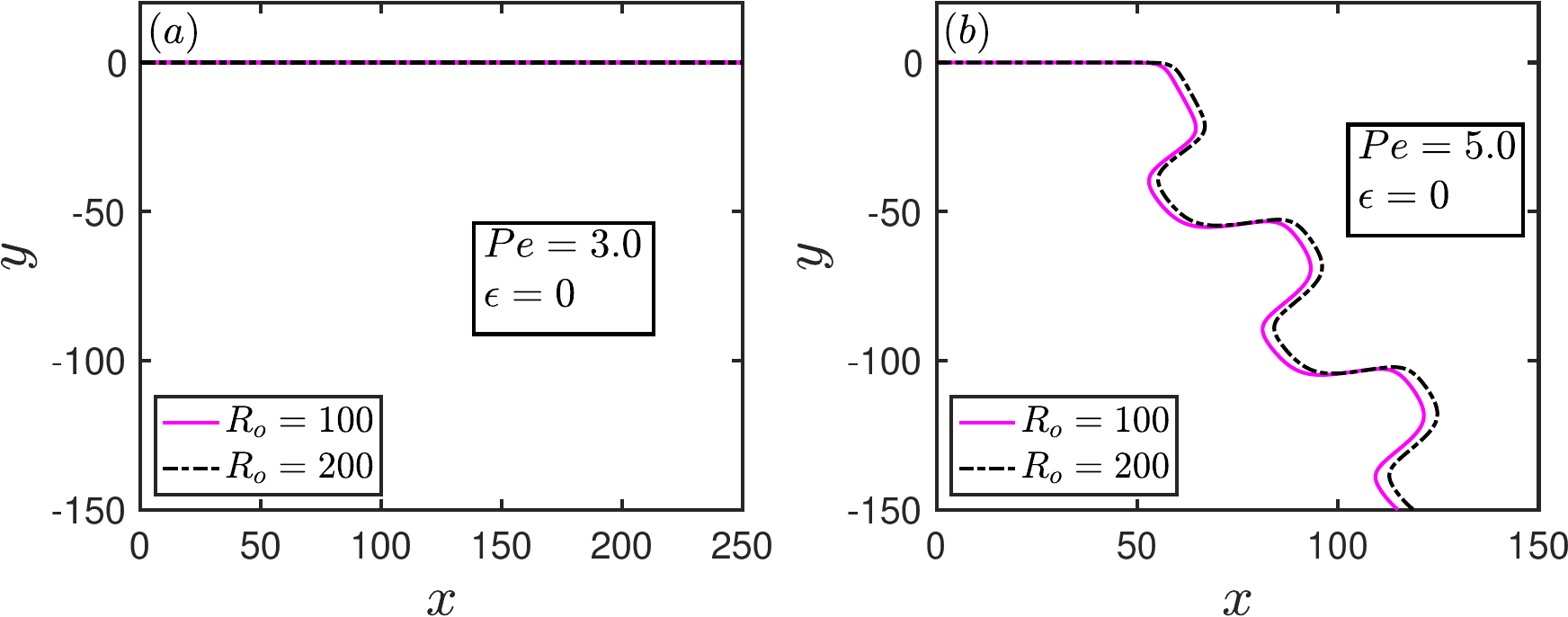}
\caption{Trajectory in the $x-y$ plane for an autophoretic disk in a quiescent fluid ($\epsilon=0$) with (a) Pe=3 and (b) Pe=5, at different values of $R_{o}$ and $\Delta t=1.0$.}
\label{fig:traj_comp}
\end{figure}

\begin{figure}[h]
\centering
\includegraphics[width=5in,height=!]{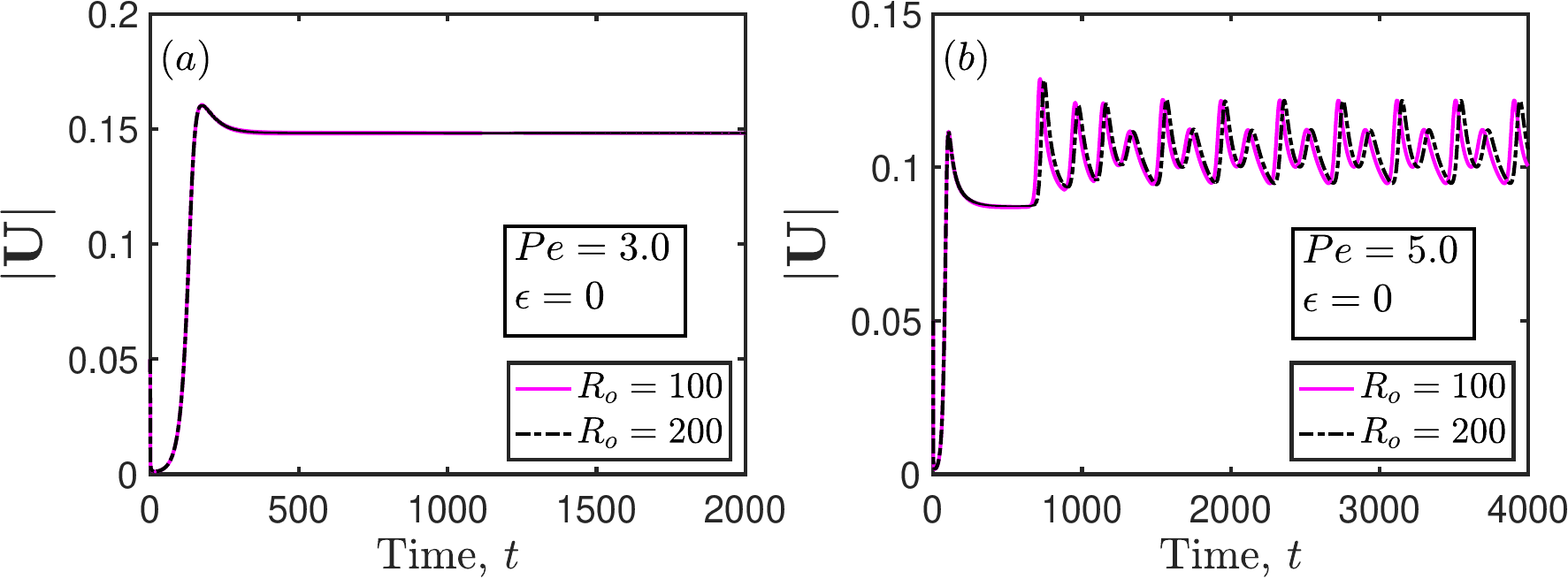}
\caption{Magnitude of the instantaneous phoretic velocity for an autophoretic disk in a quiescent fluid ($\epsilon=0$) with (a) Pe=3 and (b) Pe=5, at different values of $R_{o}$ and $\Delta t=1.0$.}
\label{fig:vmag_comp}
\end{figure}

The trajectory and phoretic velocity of an autophoretic disk in a quiescent fluid ($\epsilon=0$) at two values of the P\'{e}clet number are shown in figs.~\ref{fig:traj_comp} and~\ref{fig:vmag_comp}, respectively, computed using different values of the size of the computational domain, $R_{o}$. The good agreement between the results computed at $R_{o}=100$ and $R_{o}=200$ establishes convergence with respect to $R_{o}$. The results in figs.~\ref{fig:u_v_pe_validn} and~\ref{fig:pe20_compare_approach} were computed using $R_{o}=200$ and a timestep width of $\Delta t=1.0$, and the reasonable agreement with the predicted trends in prior work justifies the use of these parameters for the simulation of an autophoretic disk in a quiescent fluid. 

\begin{figure}
\centering
\includegraphics[width=4.5in,height=!]{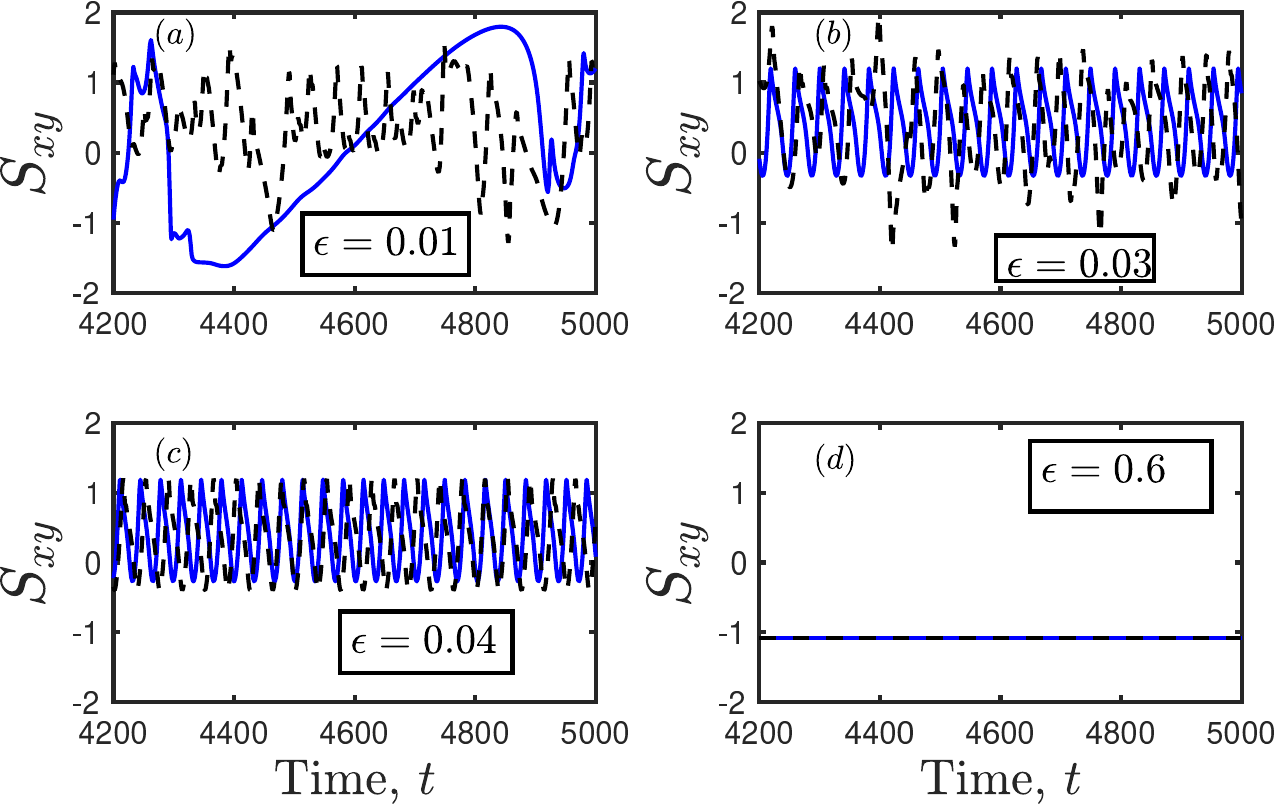}
\caption{Magnitude of the instantaneous phoretic velocity for an autophoretic disk with $Pe=20$ in an ambient shear flow, computed with $R_{o}=200$ at (a) $\epsilon=0.01$, (b) $\epsilon=0.03$, (c) $\epsilon=0.04$, and (d) $\epsilon=0.6$. The results are computed at two values of the timestep width, $\Delta t=0.1$ (\protect \blueline) and $\Delta t=1.0$ (\protect \blackline).}
\label{fig:vmag_with_sr_comp}
\end{figure}

\begin{figure}
\centering
\includegraphics[width=4.5in,height=!]{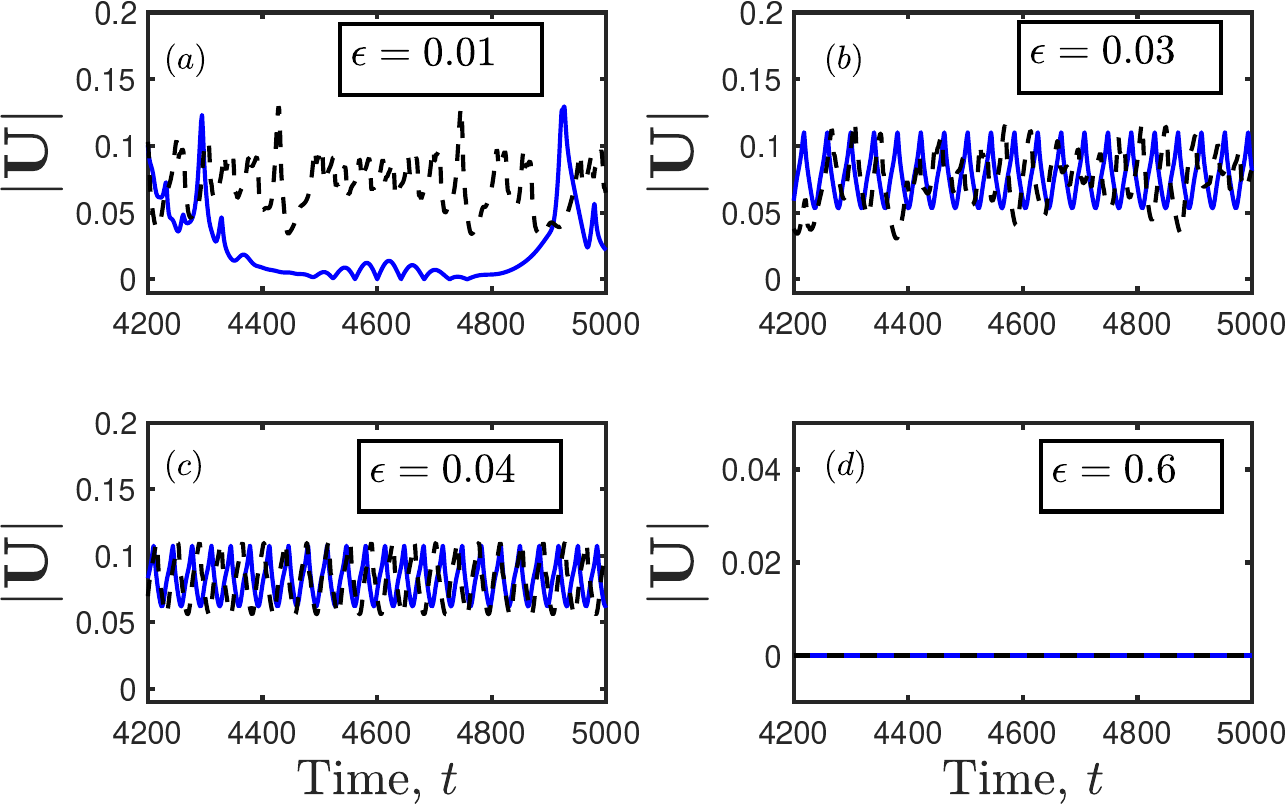}
\caption{Magnitude of the $xy$-component of the stresslet for an autophoretic disk with $Pe=20$ in an ambient shear flow, computed with $R_{o}=200$ at (a) $\epsilon=0.01$, (b) $\epsilon=0.03$, (c) $\epsilon=0.04$, and (d) $\epsilon=0.6$. The results are computed at two values of the timestep width, $\Delta t=0.1$ (\protect \blueline) and $\Delta t=1.0$ (\protect \blackline).}
\label{fig:slet_with_sr_comp}
\end{figure}

For autophoretic disk with $Pe=20$, the instantaneous phoretic velocity and the $xy$ component of the stresslet are computed at four different values of the dimensionless shear rate, at two values of the discrete timestep, $\Delta t=0.1$ and $\Delta t=1.0$, and plotted in figs.~\ref{fig:vmag_with_sr_comp}, and~\ref{fig:slet_with_sr_comp}, respectively. While the curves in all the cases do not superimpose, they follow the same qualitative trend, namely, a transition from chaotic to periodic motion, and ultimately attaining a steady state as the shear rate is increased. We therefore believe that the key results of the paper are not affected by the choice of a smaller timestep width.

\section{\label{sec:slet_exp}Expressions for the stresslet components}

Starting from the expression for the stresslet given by
\begin{equation}
\bm{S}=-2\int\left[\bm{n}\bm{v}_{\text{s}}+\bm{v}_{\text{s}}\bm{n}\right]d\theta,
\end{equation}
with
\begin{equation}
\begin{split}
\bm{n}&=\bm{e}_{r}=\cos\theta\bm{e}_{x}+\sin\theta\bm{e}_{y}\\
\bm{v}_{\text{s}}&=f\bm{e}_{\theta}=-f\sin\theta\bm{e}_{x}+f\cos\theta\bm{e}_{y}\\
f&\equiv|\bm{v}_{\text{s}}|=M\left(\cos\theta\dfrac{\partial c}{\partial y}-\sin\theta\dfrac{\partial c}{\partial x}\right),
\end{split}
\end{equation}
we may write
\begin{equation}
S_{xy}=-2\int\left[n_{x}v_{\text{s},y}+n_{y}v_{\text{s},x}\right]d\theta=-2\int f\left[\cos^2\theta-\sin^2\theta\right]d\theta=-2\int f\cos2\theta d\theta
\end{equation}
and similarly,
\begin{equation}\label{eq:s_xx_def}
S_{xx}=2\int f\sin2\theta d\theta
\end{equation}
\begin{equation}\label{eq:s_yy_def}
S_{yy}=-2\int f\sin2\theta d\theta
\end{equation}
It is clear from eqs.~(\ref{eq:s_xx_def}) and~(\ref{eq:s_yy_def}) that $S_{xx}$ and $S_{yy}$ differ only in sign, and are identical in magnitude. We have therefore chosen to report only $S_{xy}$ and $S_{yy}$ in the main paper.

\section{\label{sec:power_spec}Power spectrum analysis}

\begin{figure}[t]
\centering
\includegraphics[width=90mm]{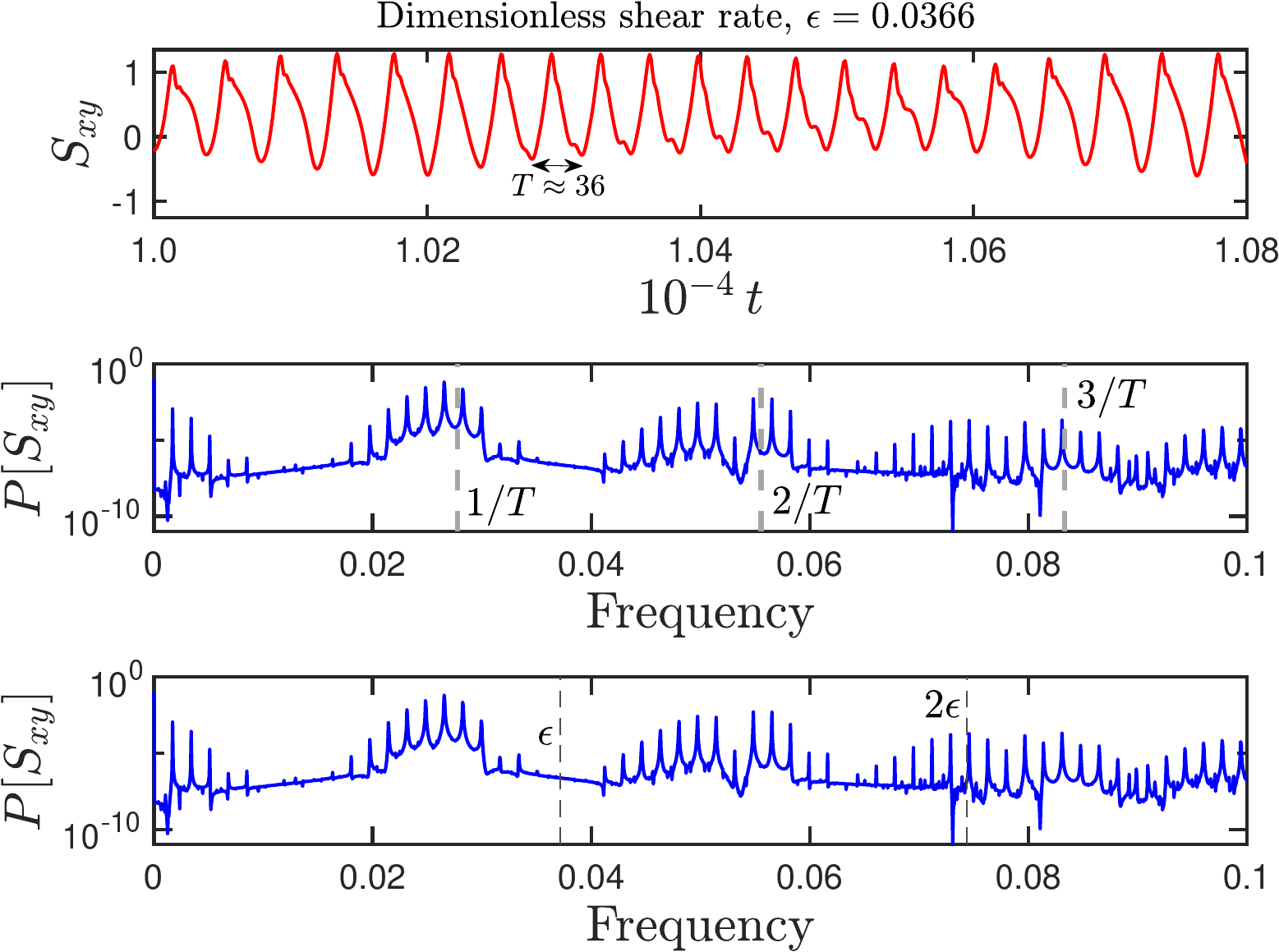}
\caption{Time series and power spectra of $S_{xy}$ at a dimensionless shear rate of $\epsilon=0.0366$.}
\label{fig:sr0_366}
\end{figure}

\begin{figure}[t]
\centering
\includegraphics[width=90mm]{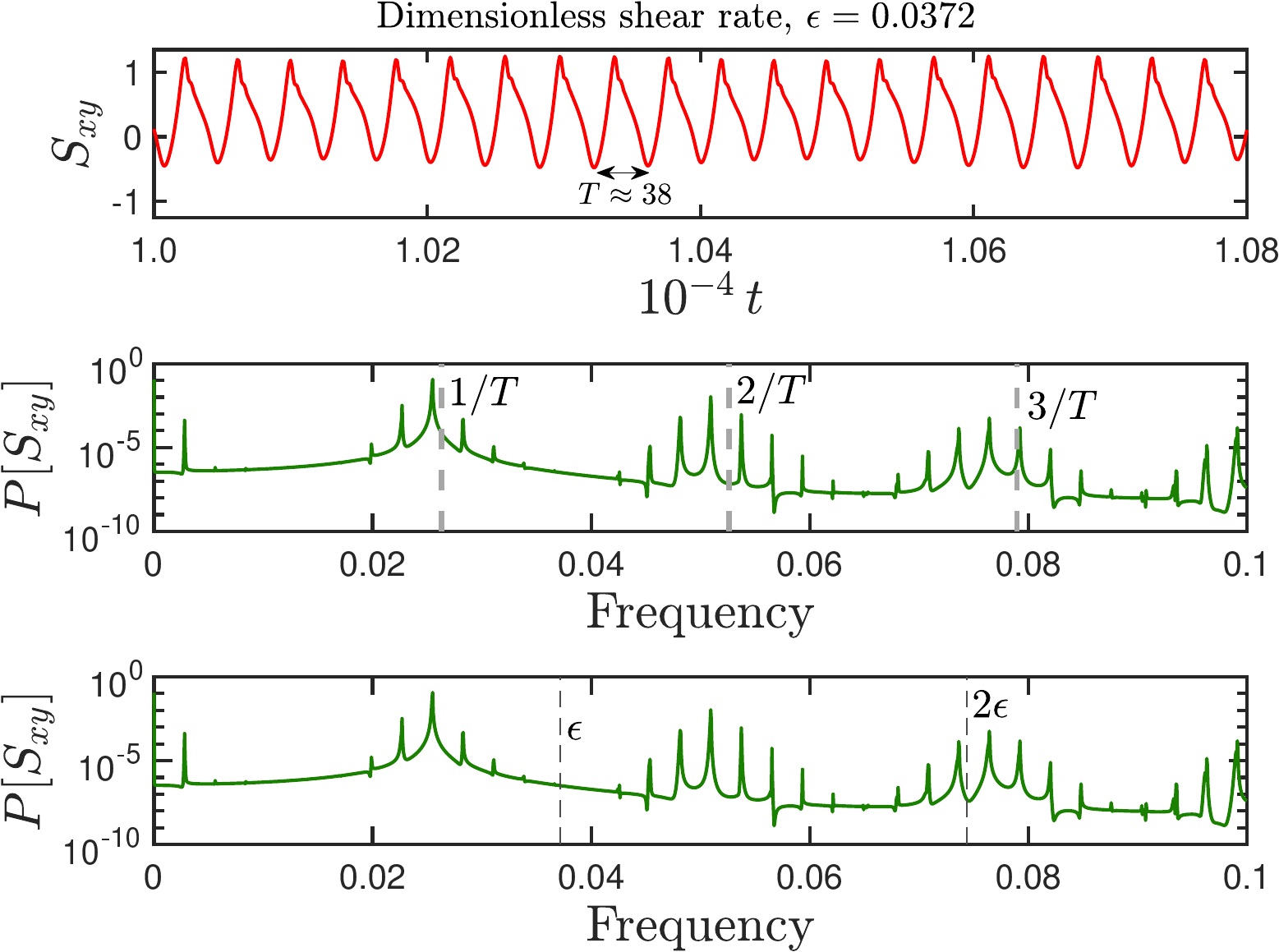}
\caption{Time series and power spectra of $S_{xy}$ at a dimensionless shear rate of $\epsilon=0.0372$.}
\label{fig:sr0_372}
\end{figure}

\begin{figure}[t]
\centering
\includegraphics[width=80mm]{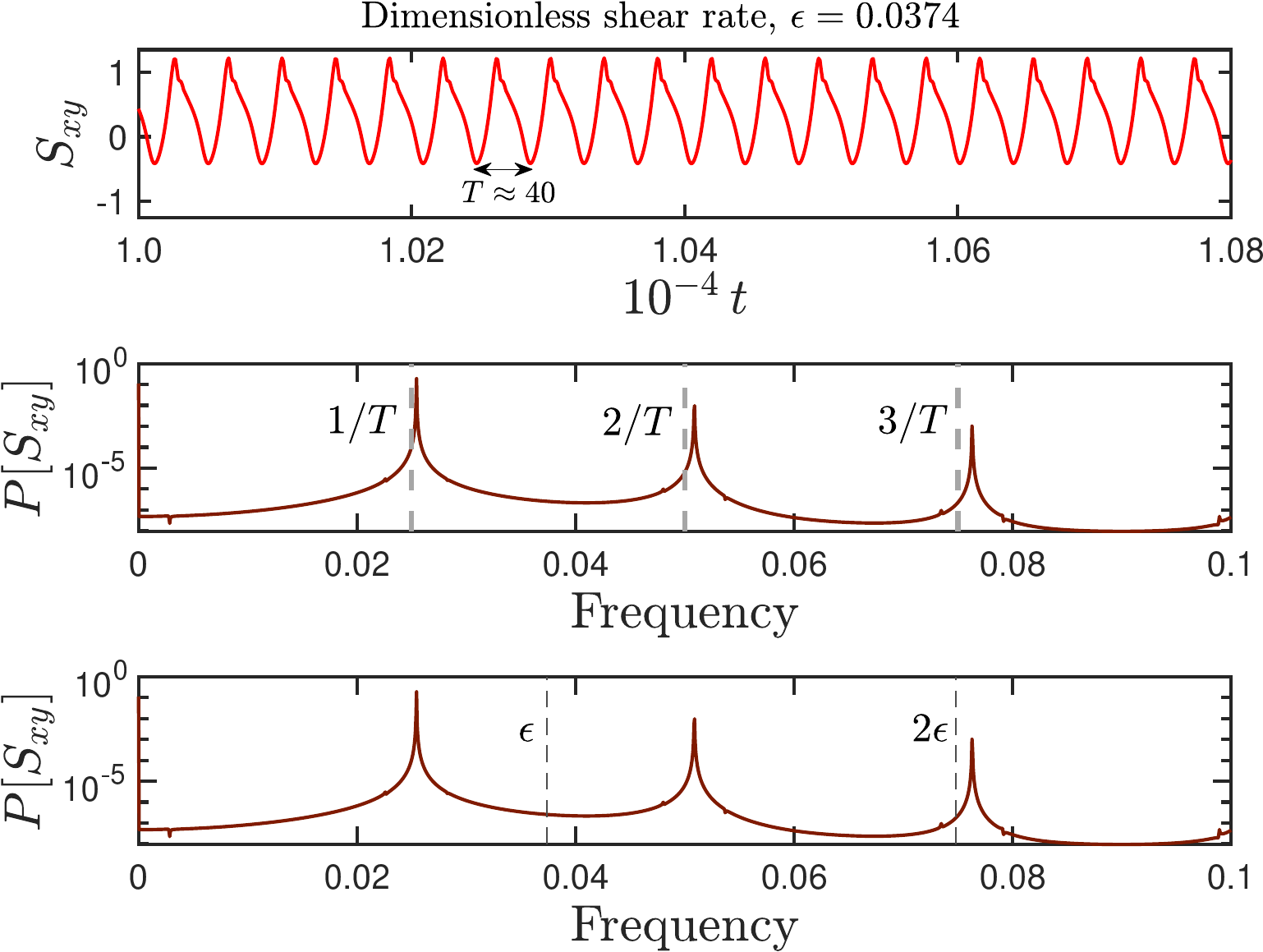}
\caption{Time series and power spectra of $S_{xy}$ at a dimensionless shear rate of $\epsilon=0.0374$.}
\label{fig:sr0_374}
\end{figure}

The only imposed frequency for an autophoretic disk in a shear flow field is the applied shear rate, $\epsilon$. It is clear from figs.~\ref{fig:sr0_366}-~\ref{fig:sr0_374}, however, that the spikes in the power spectrum do not occur at the applied shear rate or its integral multiples. A visual examination of the $S_{xy}$ time-series, however, reveals timescales that coincide more accurately with the locations of the spikes in the power spectrum. While it is not clear how these underlying frequencies may be predicted \textit{a priori}, we do note that the timescales obtained from visual inspection increases as the shear rate is increased. This trend is in agreement with the expectation that periodic variation in $S_{xy}$ disappears at high shear rates, and is replaced by a value of $S_{xy}$ that is steady in time.

\begin{figure}[t]
\centering
\includegraphics[width=80mm]{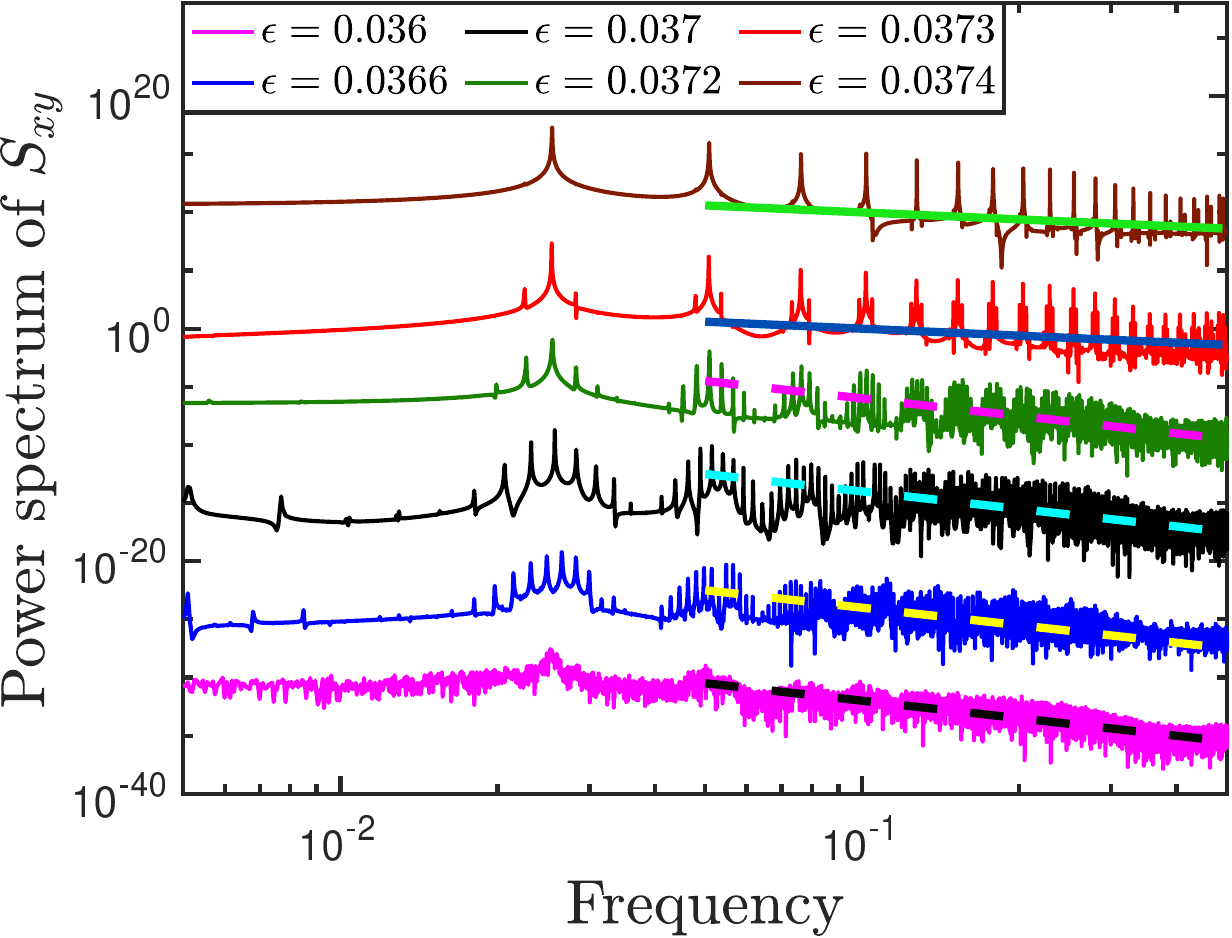}
\caption{Power spectrum of the $xy$ component of the stresslet at various values of the shear rate ($\epsilon$), and a fixed P\'{e}clet number of $Pe=20$. The ordinates of the data-series have been multiplied by a scale factor to render them well-spaced on the $y$-axis, for clarity. From bottom to top, the values of $\epsilon$ (and the associated scale factors) are 0.036 ($10^{-26}$), 0.0366($10^{-18}$), 0.037($10^{-8}$), 0.0372(1), 0.0373($10^{8}$), and 0.0374($10^{18}$). The dashed-lines indicate a slope of $-5$, while the continuous lines have a slope of $-2$.}
\label{fig:log_ax}
\end{figure}

In fig.~\ref{fig:log_ax}, Fig.~5 of the main text is replotted using logarithmic scales for both the $x$- and $y$-coordinates. The highest frequency ($\nu$) accessible in the power spectrum is limited by the discrete timestep used in the simulations ($\Delta t=1.0$ in our case). We observe a power-law scaling in the spectrum, $P[S_{xy}]\sim\nu^{n}$ with $n=-5$ for $\epsilon=\left[0.036,0.0366,0.037,0.0372\right]$, and  $n=-2$ for $\epsilon=\left[0.0373,0.0374\right]$.


\bibliography{supplement}